\input harvmac
\overfullrule=0pt
\parindent 25pt
\tolerance=10000

\input epsf

\newcount\figno
\figno=0
\def\fig#1#2#3{
\par\begingroup\parindent=0pt\leftskip=1cm\rightskip=1cm\parindent=0pt
\baselineskip=11pt
\global\advance\figno by 1
\midinsert
\epsfxsize=#3
\centerline{\epsfbox{#2}}
\vskip 12pt
{\bf Fig.\ \the\figno: } #1\par
\endinsert\endgroup\par
}
\def\figlabel#1{\xdef#1{\the\figno}}
\def\encadremath#1{\vbox{\hrule\hbox{\vrule\kern8pt\vbox{\kern8pt
\hbox{$\displaystyle #1$}\kern8pt}
\kern8pt\vrule}\hrule}}

\def\half{{\textstyle{1\over2}}}

\def\apm{{\alpha^{\prime}}}

\def\half{{1\over 2}}
 
 \def\m{{\mu}}
 \def\w{{\omega}}

 \def\t{{\theta}}
 \def\a{{\alpha}}
 
 \def\frac#1#2{{#1\over #2}}

 \def\s{{\sigma}}

 \def\Ph{{\Phi }}

 \def\p{\partial}

 \def\apm{{\alpha'}}
 \def\r{\rightarrow}

 \def\al{\alpha'}
 \def\de{\partial}

 \def\f {\frac}
 \def\ti{\tilde}

 \def\ddd{\cdot\cdot\cdot}
 
 \def\la{\langle}
 \def\lb{\rangle}

\lref\KosteleckyNT{
V.~A.~Kostelecky and S.~Samuel,
``On A Nonperturbative Vacuum For The Open Bosonic String,''
Nucl.\ Phys.\ B {\bf 336}, 263 (1990).
}

\lref\SenNF{
A.~Sen,
``Tachyon dynamics in open string theory,''
arXiv:hep-th/0410103.
}

\lref\ZwiebachIE{
B.~Zwiebach,
``Closed string field theory: Quantum action and the B-V master equation,''
Nucl.\ Phys.\ B {\bf 390}, 33 (1993)
[arXiv:hep-th/9206084].
}

\lref\PolchinskiRR{
J.~Polchinski,
``String Theory. Vol. 2: Superstring Theory And Beyond,''
Cambridge, UK: Univ. Pr. (1998) 531 p.
}

\lref\SenZM{
A.~Sen,
 ``Rolling tachyon boundary state, conserved charges and two 
dimensional string theory,''
arXiv:hep-th/0402157.
}

\lref\AdPoSi{
A.~Adams, J.~Polchinski and E.~Silverstein,
``Don't panic! Closed string tachyons in ALE space-times,''
JHEP {\bf 0110} (2001) 029
[arXiv:hep-th/0108075].}

\lref\Dabh{
A.~Dabholkar,
``Tachyon condensation and black hole entropy,''
arXiv:hep-th/0111004.
}

\lref\DabholkarAI{
A.~Dabholkar,
``Strings on a cone and black hole entropy,''
Nucl.\ Phys.\ B {\bf 439}, 650 (1995)
[arXiv:hep-th/9408098].
}

\lref\LoweAH{
D.~A.~Lowe and A.~Strominger,
``Strings near a Rindler or black hole horizon,''
Phys.\ Rev.\ D {\bf 51}, 1793 (1995)
[arXiv:hep-th/9410215].
}

\lref\DoKaPoSh{
M.~R.~Douglas, D.~Kabat, P.~Pouliot and S.~H.~Shenker,
``D-branes and short distances in string theory,''
Nucl.\ Phys.\ B {\bf 485}, 85 (1997)
[arXiv:hep-th/9608024].
}

\lref\chicago{ J.~A.~Harvey, D.~Kutasov, E.~J.~Martinec, G.~Moore,
``Localized tachyons and RG flows'', arXiv:hep-th/0111154.
}

\lref\Va{
C.~Vafa,
``Mirror symmetry and closed string tachyon condensation,''
[arXiv:hep-th/0111051].}

\lref\DhVa{
A.~Dabholkar and C.~Vafa,
``tt* geometry and closed string tachyon potential,''
JHEP {\bf 0202} (2002) 008
[arXiv:hep-th/0111155].}

\lref\MiTa{
S.~Minwalla and T.~Takayanagi,
``Evolution of D-branes under closed string tachyon condensation,''
JHEP {\bf 0309}, 011 (2003)
[arXiv:hep-th/0307248].
}

\lref\DaGuHeMi{
J.~R.~David, M.~Gutperle, M.~Headrick and S.~Minwalla,
``Closed string tachyon condensation on twisted circles,''
JHEP {\bf 0202}, 041 (2002)
[arXiv:hep-th/0111212].
}

\lref\GuHeMiSc{
M.~Gutperle, M.~Headrick, S.~Minwalla and V.~Schomerus,
``Space-time energy decreases under world-sheet RG flow,''
arXiv:hep-th/0211063.}

\lref\HoVa{
K.~Hori and C.~Vafa,
``Mirror symmetry,''
arXiv:hep-th/0002222.}

\lref\HoIqVa{
K.~Hori, A.~Iqbal and C.~Vafa,
``D-branes and mirror symmetry,''
arXiv:hep-th/0005247.}

\lref\HoKa{
K.~Hori and A.~Kapustin,
``Duality of the fermionic 2d black hole and N = 2 Liouville
theory as  mirror symmetry,''
JHEP {\bf 0108} (2001) 045
[arXiv:hep-th/0104202].}

\lref\BSFT{
A.~A.~Gerasimov and S.~L.~Shatashvili,
``On exact tachyon potential in open string field theory,''
JHEP {\bf 0010} (2000) 034
[arXiv:hep-th/0009103];
D.~Kutasov, M.~Marino and G.~W.~Moore,
``Some exact results on tachyon condensation in string field theory,''
JHEP {\bf 0010} (2000) 045
[arXiv:hep-th/0009148];
D.~Kutasov, M.~Marino and G.~W.~Moore,
``Remarks on tachyon condensation in superstring field theory,''
arXiv:hep-th/0010108.}

\lref\ta{
T.~Takayanagi,
``Holomorphic tachyons and fractional D-branes,''
Nucl.\ Phys.\ B {\bf 603} (2001) 259
[arXiv:hep-th/0103021];
T.~Takayanagi,
``Tachyon condensation on orbifolds and McKay correspondence,''
Phys.\ Lett.\ B {\bf 519} (2001) 137
[arXiv:hep-th/0106142].}
\lref\Ue{
T.~Uesugi,
``Worldsheet description of tachyon condensation in open string theory,''
arXiv:hep-th/0302125.
}
\lref\DoMo{
M.~R.~Douglas and G.~W.~Moore,
``D-branes, Quivers, and ALE Instantons,''
arXiv:hep-th/9603167.}

\lref\DiDoGo{
D.~E.~Diaconescu, M.~R.~Douglas and J.~Gomis,
``Fractional branes and wrapped branes,''
JHEP {\bf 9802} (1998) 013
[arXiv:hep-th/9712230].}

\lref\TaUeo{
T.~Takayanagi and T.~Uesugi,
``Orbifolds as Melvin geometry,''
JHEP {\bf 0112} (2001) 004
[arXiv:hep-th/0110099].}

\lref\TaUet{
T.~Takayanagi and T.~Uesugi,
``D-branes in Melvin background,''
JHEP {\bf 0111} (2001) 036
[arXiv:hep-th/0110200];
T.~Takayanagi and T.~Uesugi,
``Flux stabilization of D-branes in NSNS Melvin background,''
Phys.\ Lett.\ B {\bf 528} (2002) 156
[arXiv:hep-th/0112199].}

\lref\DuMo{
E.~Dudas and J.~Mourad,
``D-branes in string theory Melvin backgrounds,''
Nucl.\ Phys.\ B {\bf 622} (2002) 46
[arXiv:hep-th/0110186].}

\lref\CoGu{
M.~S.~Costa and M.~Gutperle,
``The Kaluza-Klein Melvin solution in M-theory,''
JHEP {\bf 0103} (2001) 027
[arXiv:hep-th/0012072].}

\lref\GuSt{
M.~Gutperle and A.~Strominger,
``Fluxbranes in string theory,''
JHEP {\bf 0106} (2001) 035
[arXiv:hep-th/0104136].}

\lref\Gu{
M.~Gutperle,
``A note on perturbative and nonperturbative instabilities of
twisted  circles,''
Phys.\ Lett.\ B {\bf 545}, 379 (2002)
[arXiv:hep-th/0207131].
}
\lref\RussoTF{
J.~G.~Russo and A.~A.~Tseytlin,
 ``Magnetic backgrounds and tachyonic instabilities in closed superstring
theory and M-theory,''
Nucl.\ Phys.\ B {\bf 611}, 93 (2001)
[arXiv:hep-th/0104238].
}

\lref\ZamolodchikovGT{ A.~B.~Zamolodchikov, ``'Irreversibility' of
the flux of the renormalization group in a 2-D field theory,''
JETP Lett.\  {\bf 43}, 730 (1986) [Pisma Zh.\ Eksp.\ Teor.\ Fiz.\
{\bf 43}, 565 (1986)].
}

\lref\WittenYC{
E.~Witten,
``Phases of N = 2 theories in two dimensions,''
Nucl.\ Phys.\ B {\bf 403}, 159 (1993)
[arXiv:hep-th/9301042].
}

\lref\SuyamaGD{
T.~Suyama,
``Properties of string theory on Kaluza-Klein Melvin background,''
JHEP {\bf 0207}, 015 (2002)
[arXiv:hep-th/0110077].
}

\lref\TseytlinZV{
A.~A.~Tseytlin,
``Closed superstrings in magnetic flux tube background,''
Nucl.\ Phys.\ Proc.\ Suppl.\  {\bf 49}, 338 (1996)
[arXiv:hep-th/9510041].
}

\lref\TseytlinEI{
A.~A.~Tseytlin,
``Melvin solution in string theory,''
Phys.\ Lett.\ B {\bf 346}, 55 (1995)
[arXiv:hep-th/9411198].
}

\lref\wittenbh{
E.~Witten,
``On string theory and black holes,''
Phys.\ Rev.\ D {\bf 44}, 314 (1991).
}

\lref\tse{
J.~G.~Russo and A.~A.~Tseytlin,
``Magnetic flux tube models in superstring theory,''
Nucl.\ Phys.\ B {\bf 461}, 131 (1996)
[arXiv:hep-th/9508068].
}

\lref\piljin{
Y.~Michishita and P.~Yi,
``D-brane probe and closed string tachyons,''
Phys.\ Rev.\ D {\bf 65}, 086006 (2002)
[arXiv:hep-th/0111199].
}

\lref\MaMo{
E.~J.~Martinec and G.~Moore,
``On decay of K-theory,''
arXiv:hep-th/0212059.
}

\lref\Gob{
S.~Govindarajan, T.~Jayaraman and T.~Sarkar,
``On D-branes from gauged linear sigma models,''
Nucl.\ Phys.\ B {\bf 593}, 155 (2001)
[arXiv:hep-th/0007075].
}
\lref\Vaq{
C.~Vafa,
``Quantum Symmetries Of String Vacua,''
Mod.\ Phys.\ Lett.\ A {\bf 4}, 1615 (1989).
}

\lref\Bi{
M.~Billo, B.~Craps and F.~Roose,
``Orbifold boundary states from Cardy's condition,''
JHEP {\bf 0101}, 038 (2001)
[arXiv:hep-th/0011060].
}
\lref\Martinec{
E.~J.~Martinec,
``Defects, decay, and dissipated states,''
arXiv:hep-th/0210231.
}


\lref\SenBA{
A.~Sen,
``Tachyon condensation on the brane antibrane system,''
JHEP {\bf 9808}, 012 (1998)
[arXiv:hep-th/9805170].
}

\lref\SenNB{
A.~Sen,
``SO(32) spinors of type I and other solitons on brane-antibrane pair,''
JHEP {\bf 9809}, 023 (1998)
[arXiv:hep-th/9808141].
}

\lref\SenBO{
A.~Sen,
``Descent relations among bosonic D-branes,''
Int.\ J.\ Mod.\ Phys.\ A {\bf 14}, 4061 (1999)
[arXiv:hep-th/9902105].
}

\lref\SenRe{
A.~Sen,
``Non-BPS states and branes in string theory,''
arXiv:hep-th/9904207.
}

\lref\BG{
O.~Bergman and M.~R.~Gaberdiel,
``Non-BPS Dirichlet branes,''
Class.\ Quant.\ Grav.\  {\bf 17}, 961 (2000)
[arXiv:hep-th/9908126].
}

\lref\LR{
A.~Lerda and R.~Russo,
``Stable non-BPS states in string theory: A pedagogical review,''
Int.\ J.\ Mod.\ Phys.\ A {\bf 15}, 771 (2000)
[arXiv:hep-th/9905006].
}


\lref\Oh{
K.~Ohmori,
``A review on tachyon condensation in open string field theories,''
arXiv:hep-th/0102085.
}

\lref\DeS{ P.~J.~De Smet,
``Tachyon condensation: Calculations in
string field theory,'' arXiv:hep-th/0109182.
}

\lref\ABG{ I.~Y.~Arefeva, D.~M.~Belov, A.~A.~Giryavets,
A.~S.~Koshelev and P.~B.~Medvedev, ``Noncommutative field theories
and (super)string field theories,'' arXiv:hep-th/0111208.
}

\lref\BonoraXP{ L.~Bonora, C.~Maccaferri, D.~Mamone and
M.~Salizzoni, ``Topics in string field theory,''
arXiv:hep-th/0304270.
}

\lref\WiSCFT{
E.~Witten,
``Noncommutative Geometry And String Field Theory,''
Nucl.\ Phys.\ B {\bf 268}, 253 (1986).
}

\lref\SenUN{
A.~Sen,
``Universality of the tachyon potential,''
JHEP {\bf 9912}, 027 (1999)
[arXiv:hep-th/9911116].
}

\lref\SenCS{
A.~Sen and B.~Zwiebach,
``Tachyon condensation in string field theory,''
JHEP {\bf 0003}, 002 (2000)
[arXiv:hep-th/9912249].
}

\lref\MT{
N.~Moeller and W.~Taylor,
``Level truncation and the tachyon in open bosonic string field theory,''
Nucl.\ Phys.\ B {\bf 583}, 105 (2000)
[arXiv:hep-th/0002237].
}

\lref\SenBK{
N.~Berkovits, A.~Sen and B.~Zwiebach,
``Tachyon condensation in superstring field theory,''
Nucl.\ Phys.\ B {\bf 587}, 147 (2000)
[arXiv:hep-th/0002211].
}

\lref\HK{
J.~A.~Harvey and P.~Kraus,
``D-branes as unstable lumps in bosonic open string field theory,''
JHEP {\bf 0004}, 012 (2000)
[arXiv:hep-th/0002117].
}

\lref\MJMT{ R.~de Mello Koch, A.~Jevicki, M.~Mihailescu and
R.~Tatar, ``Lumps and p-branes in open string field theory,''
Phys.\ Lett.\ B {\bf 482}, 249 (2000) [arXiv:hep-th/0003031].
}

\lref\MSZ{
N.~Moeller, A.~Sen and B.~Zwiebach,
``D-branes as tachyon lumps in string field theory,''
JHEP {\bf 0008}, 039 (2000)
[arXiv:hep-th/0005036].
}

\lref\TaE{
W.~Taylor,
``D-brane effective field theory from string field theory,''
Nucl.\ Phys.\ B {\bf 585}, 171 (2000)
[arXiv:hep-th/0001201].
}

\lref\TZRE{
W.~Taylor and B.~Zwiebach,
``D-branes, tachyons, and string field theory,''
arXiv:hep-th/0311017.
}

\lref\TaRE{
W.~Taylor,
``Lectures on D-branes, tachyon condensation, and string field theory,''
arXiv:hep-th/0301094.
}

\lref\MTL{
N.~Moeller and W.~Taylor,
``Level truncation and the tachyon in open bosonic string field theory,''
Nucl.\ Phys.\ B {\bf 583}, 105 (2000)
[arXiv:hep-th/0002237].
}

\lref\BeSFT{
N.~Berkovits,
``SuperPoincare invariant superstring field theory,''
Nucl.\ Phys.\ B {\bf 450}, 90 (1995)
[Erratum-ibid.\ B {\bf 459}, 439 (1996)]
[arXiv:hep-th/9503099];
``The Ramond sector of open superstring field theory,''
JHEP {\bf 0111}, 047 (2001)
[arXiv:hep-th/0109100].
}

\lref\BeRe{
N.~Berkovits,
``Review of open superstring field theory,''
arXiv:hep-th/0105230.
}

\lref\GRE{
D.~Gaiotto and L.~Rastelli,
``Experimental string field theory,''
JHEP {\bf 0308}, 048 (2003)
[arXiv:hep-th/0211012].
}


\lref\HKM{
J.~A.~Harvey, D.~Kutasov and E.~J.~Martinec,
``On the relevance of tachyons,''
arXiv:hep-th/0003101.
}

\lref\AL{
I.~Affleck and A.~W.~W.~Ludwig,
``Universal Noninteger 'Ground State Degeneracy' In Critical Quantum
Systems,''
Phys.\ Rev.\ Lett.\  {\bf 67}, 161 (1991).
}
\lref\KO{
A.~Konechny,
``g-function in perturbation theory,''
arXiv:hep-th/0310258.
}
\lref\HKMS{
J.~A.~Harvey, S.~Kachru, G.~W.~Moore and E.~Silverstein,
``Tension is dimension,''
JHEP {\bf 0003}, 001 (2000)
[arXiv:hep-th/9909072].
}

\lref\WiBSFT{
E.~Witten,
``On background independent open string field theory,''
Phys.\ Rev.\ D {\bf 46}, 5467 (1992)
[arXiv:hep-th/9208027];
E.~Witten,
``Some computations in background independent off-shell string theory,''
Phys.\ Rev.\ D {\bf 47}, 3405 (1993)
[arXiv:hep-th/9210065].
}
\lref\TsB{
A.~A.~Tseytlin,
``Sigma model approach to string theory effective actions with tachyons,''
J.\ Math.\ Phys.\  {\bf 42}, 2854 (2001)
[arXiv:hep-th/0011033].
}
\lref\Sh{
S.~L.~Shatashvili,
``Comment on the background independent open string theory,''
Phys.\ Lett.\ B {\bf 311}, 83 (1993)
[arXiv:hep-th/9303143];
S.~L.~Shatashvili,
``On the problems with background independence in string theory,''
arXiv:hep-th/9311177.
}

\lref\GeSh{
A.~A.~Gerasimov and S.~L.~Shatashvili,
``On exact tachyon potential in open string field theory,''
JHEP {\bf 0010}, 034 (2000)
[arXiv:hep-th/0009103];
A.~A.~Gerasimov and S.~L.~Shatashvili,
``Stringy Higgs mechanism and the fate of open strings,''
JHEP {\bf 0101}, 019 (2001)
[arXiv:hep-th/0011009].
}

\lref\KMMB{ D.~Kutasov, M.~Marino and G.~W.~Moore, ``Remarks on
tachyon condensation in superstring field theory,''
arXiv:hep-th/0010108.
}

\lref\KMMS{
D.~Kutasov, M.~Marino and G.~W.~Moore,
``Some exact results on tachyon condensation in string field theory,''
JHEP {\bf 0010}, 045 (2000)
[arXiv:hep-th/0009148].
}

\lref\BSFTBA{
K.~Hori,
``Linear models of supersymmetric D-branes,''
arXiv:hep-th/0012179;
P.~Kraus and F.~Larsen,
``Boundary string field theory of the DD-bar system,''
Phys.\ Rev.\ D {\bf 63}, 106004 (2001)
[arXiv:hep-th/0012198];
T.~Takayanagi, S.~Terashima and T.~Uesugi,
``Brane-antibrane action from boundary string field theory,''
JHEP {\bf 0103}, 019 (2001)
[arXiv:hep-th/0012210].
}

\lref\BSFTBV{
M.~Marino,
``On the BV formulation of boundary superstring field theory,''
JHEP {\bf 0106}, 059 (2001)
[arXiv:hep-th/0103089];
V.~Niarchos and N.~Prezas,
``Boundary superstring field theory,''
Nucl.\ Phys.\ B {\bf 619}, 51 (2001)
[arXiv:hep-th/0103102].
}


\lref\GSS{
M.~Gutperle and A.~Strominger,
``Spacelike branes,''
JHEP {\bf 0204}, 018 (2002)
[arXiv:hep-th/0202210].
}

\lref\CKLM{
C.~G.~Callan, I.~R.~Klebanov, A.~W.~W.~Ludwig and J.~M.~Maldacena,
``Exact solution of a boundary conformal field theory,''
Nucl.\ Phys.\ B {\bf 422}, 417 (1994)
[arXiv:hep-th/9402113];
J.~Polchinski and L.~Thorlacius,
``Free Fermion Representation Of A Boundary Conformal Field Theory,''
Phys.\ Rev.\ D {\bf 50}, 622 (1994)
[arXiv:hep-th/9404008].
}

\lref\RSM{
A.~Recknagel and V.~Schomerus,
``Boundary deformation theory and moduli spaces of D-branes,''
Nucl.\ Phys.\ B {\bf 545}, 233 (1999)
[arXiv:hep-th/9811237].
}

\lref\OS{
T.~Okuda and S.~Sugimoto,
``Coupling of rolling tachyon to closed strings,''
Nucl.\ Phys.\ B {\bf 647}, 101 (2002)
[arXiv:hep-th/0208196].
}

\lref\CLL{
B.~Chen, M.~Li and F.~L.~Lin,
``Gravitational radiation of rolling tachyon,''
JHEP {\bf 0211}, 050 (2002)
[arXiv:hep-th/0209222].
}

\lref\LLM{
N.~Lambert, H.~Liu and J.~Maldacena,
``Closed strings from decaying D-branes,''
arXiv:hep-th/0303139.
}

\lref\KLMS{
J.~L.~Karczmarek, H.~Liu, J.~Maldacena and A.~Strominger,
``UV finite brane decay,''
JHEP {\bf 0311}, 042 (2003)
[arXiv:hep-th/0306132].
}

\lref\SOP{
A.~Strominger,
``Open string creation by S-branes,''
arXiv:hep-th/0209090.
}

\lref\GSL{
M.~Gutperle and A.~Strominger,
``Timelike boundary Liouville theory,''
Phys.\ Rev.\ D {\bf 67}, 126002 (2003)
[arXiv:hep-th/0301038].
}

\lref\MSY{
A.~Maloney, A.~Strominger and X.~Yin,
``S-brane thermodynamics,''
JHEP {\bf 0310}, 048 (2003)
[arXiv:hep-th/0302146].
}

\lref\LNT{
F.~Larsen, A.~Naqvi and S.~Terashima,
``Rolling tachyons and decaying branes,''
JHEP {\bf 0302}, 039 (2003)
[arXiv:hep-th/0212248].
}

\lref\IqbalST{
A.~Iqbal and A.~Naqvi,
``Tachyon condensation on a non-BPS D-brane,''
arXiv:hep-th/0004015.
}

\lref\DeSmetJE{
P.~J.~De Smet and J.~Raeymaekers,
``The tachyon potential in Witten's superstring field theory,''
JHEP {\bf 0008}, 020 (2000)
[arXiv:hep-th/0004112].
}

\lref\SenRO{
A.~Sen,
``Rolling tachyon,''
JHEP {\bf 0204}, 048 (2002)
[arXiv:hep-th/0203211].
}

\lref\SenTM{
A.~Sen,
``Tachyon matter,''
JHEP {\bf 0207}, 065 (2002)
[arXiv:hep-th/0203265].
}

\lref\SenFT{
A.~Sen,
``Field theory of tachyon matter,''
Mod.\ Phys.\ Lett.\ A {\bf 17}, 1797 (2002)
[arXiv:hep-th/0204143].
}

\lref\SenOC{
A.~Sen,
``Open and closed strings from unstable D-branes,''
Phys.\ Rev.\ D {\bf 68}, 106003 (2003)
[arXiv:hep-th/0305011].
}

\lref\SenDU{
A.~Sen,
``Open-closed duality at tree level,''
Phys.\ Rev.\ Lett.\  {\bf 91}, 181601 (2003)
[arXiv:hep-th/0306137].
}

\lref\GRB{
D.~Gaiotto, N.~Itzhaki and L.~Rastelli,
``Closed strings as imaginary D-branes,''
arXiv:hep-th/0304192.
}

\lref\BeT{N.~Berkovits, ``The tachyon potential in open
Neveu-Schwarz string field theory,'' JHEP {\bf 0004}, 022 (2000)
[arXiv:hep-th/0001084].
}

\lref\He{
M.~Headrick,
``Decay of ${\bf C}/{\bf Z}_n$: Exact supergravity solutions,''
JHEP {\bf 0403}, 025 (2004)
[arXiv:hep-th/0312213].
}

\lref\GrHa{
R.~Gregory and J.~A.~Harvey,
``Spacetime decay of cones at strong coupling,''
Class.\ Quant.\ Grav.\  {\bf 20}, L231 (2003)
[arXiv:hep-th/0306146].
}

\lref\OkZw{
Y. Okawa and B. Zwiebach,
``Twisted tachyon condensation in closed string field theory,"
arXiv:hep-th/0403051.
}

\lref\DaIqRa{
A.~Dabholkar, A.~Iqubal and J.~Raeymaekers,
``Off-shell interactions for closed-string tachyons,''
arXiv:hep-th/0403238.
}

\lref\WiBON{
E.~Witten,
``Instability Of The Kaluza-Klein Vacuum,''
Nucl.\ Phys.\ B {\bf 195}, 481 (1982).
}

\lref\EmGu{
R.~Emparan and M.~Gutperle,
``From p-branes to fluxbranes and back,''
JHEP {\bf 0112}, 023 (2001)
[arXiv:hep-th/0111177].
}
\lref\MoP{
G.~Moore and A.~Parnachev,
``Localized Tachyons and the Quantum McKay Correspondence,''
arXiv:hep-th/0403016.
}



\lref\NamTS{
S.~k.~Nam and S.~J.~Sin,
``Condensation of localized tachyons and spacetime supersymmetry,''
J.\ Korean Phys.\ Soc.\  {\bf 43}, 34 (2003)
[arXiv:hep-th/0201132].
}

\lref\SinQS{
S.~J.~Sin,
``Tachyon mass, c-function and counting localized degrees of freedom,''
Nucl.\ Phys.\ B {\bf 637}, 395 (2002)
[arXiv:hep-th/0202097].
}

\lref\BasuJT{
A.~Basu,
``Localized tachyons and the g(cl) conjecture,''
JHEP {\bf 0207}, 011 (2002)
[arXiv:hep-th/0204247].
}

\lref\ChaudhuriJJ{
S.~Chaudhuri,
``Thermal instabilities and the g-theorem,''
arXiv:hep-th/0212220.
}

\lref\SinXE{
S.~J.~Sin,
``Localized tachyon mass and a g-theorem analogue,''
Nucl.\ Phys.\ B {\bf 667}, 310 (2003)
[arXiv:hep-th/0308015].
}


\lref\KrausCB{
P.~Kraus, A.~Ryzhov and M.~Shigemori,
``Strings in noncompact spacetimes: Boundary terms and conserved
charges,''
Phys.\ Rev.\ D {\bf 66}, 106001 (2002)
[arXiv:hep-th/0206080].
}

\lref\DavidKM{
J.~R.~David,
``Unstable magnetic fluxes in heterotic string theory,''
JHEP {\bf 0209}, 006 (2002)
[arXiv:hep-th/0208011].
}


\lref\DaCunhaFM{
B.~C.~Da Cunha and E.~J.~Martinec,
``Closed string tachyon condensation and worldsheet inflation,''
Phys.\ Rev.\ D {\bf 68}, 063502 (2003)
[arXiv:hep-th/0303087].
}

\lref\StromingerFN{
A.~Strominger and T.~Takayanagi,
``Correlators in timelike bulk Liouville theory,''
Adv.\ Theor.\ Math.\ Phys.\  {\bf 7}, 369 (2003)
[arXiv:hep-th/0303221].
}

\lref\SchomerusVV{
V.~Schomerus,
``Rolling tachyons from Liouville theory,''
JHEP {\bf 0311}, 043 (2003)
[arXiv:hep-th/0306026].
}


\lref\CostaNW{
M.~S.~Costa and M.~Gutperle,
``The Kaluza-Klein Melvin solution in M-theory,''
JHEP {\bf 0103}, 027 (2001)
[arXiv:hep-th/0012072].
}


\lref\MaggioreQR{
M.~Maggiore,
 ``The Atick-Witten free energy, closed tachyon condensation and deformed
Poincare symmetry,''
Nucl.\ Phys.\ B {\bf 647}, 69 (2002)
[arXiv:hep-th/0205014].
}


\lref\BardakciGV{
K.~Bardakci,
``Dual Models And Spontaneous Symmetry Breaking,''
Nucl.\ Phys.\ B {\bf 68}, 331 (1974).
}

\lref\BardakciUI{
K.~Bardakci,
``Dual Models And Spontaneous Symmetry Breaking II,''
Nucl.\ Phys.\ B {\bf 70}, 397 (1974).
}

\lref\BardakciVS{
K.~Bardakci and M.~B.~Halpern,
``Explicit Spontaneous Breakdown In A Dual Model,''
Phys.\ Rev.\ D {\bf 10}, 4230 (1974).
}

\lref\BardakciUX{
K.~Bardakci and M.~B.~Halpern,
``Explicit Spontaneous Breakdown In A Dual Model. 2. N Point Functions,''
Nucl.\ Phys.\ B {\bf 96}, 285 (1975).
}

\lref\BardakciAN{
K.~Bardakci,
``Spontaneous Symmetry Breakdown In The Standard Dual String Model,''
Nucl.\ Phys.\ B {\bf 133}, 297 (1978).
}

\lref\SuyamaXK{
T.~Suyama,
``Deformation of CHS model,''
Nucl.\ Phys.\ B {\bf 641}, 341 (2002)
[arXiv:hep-th/0206171].
}

\lref\SuyamaKY{
T.~Suyama,
``Closed string tachyons and RG flows,''
JHEP {\bf 0210}, 051 (2002)
[arXiv:hep-th/0210054].
}

\lref\SuyamaXJ{
T.~Suyama,
``Target space approach to closed string tachyons,''
arXiv:hep-th/0302010.
}

\lref\DineCA{
M.~Dine, E.~Gorbatov, I.~R.~Klebanov and M.~Krasnitz,
``Closed string tachyons and their implications for non-supersymmetric strings,''
arXiv:hep-th/0303076.
}

\lref\SuyamaAS{
T.~Suyama,
``On decay of bulk tachyons,''
arXiv:hep-th/0308030.
}

\lref\KarczmarekPV{
J.~L.~Karczmarek and A.~Strominger,
``Matrix cosmology,''
arXiv:hep-th/0309138.
}

\lref\JoannaAndy{
J.~L.~Karczmarek and A.~Strominger,
``Closed string tachyon condensation at c = 1,''
arXiv:hep-th/0403169.
}

\lref\DavidNQ{
J.~R.~David, S.~Minwalla and C.~Nunez,
``Fermions in bosonic string theories,''
JHEP {\bf 0109}, 001 (2001)
[arXiv:hep-th/0107165].
}


\lref\SarkarDC{
S.~Sarkar and B.~Sathiapalan,
``Closed string tachyons on C/Z(N),''
arXiv:hep-th/0309029.
}

\lref\AngelantonjID{
C.~Angelantonj, E.~Dudas and J.~Mourad,
``Orientifolds of string theory Melvin backgrounds,''
Nucl.\ Phys.\ B {\bf 637}, 59 (2002)
[arXiv:hep-th/0205096].
}


\lref\LiuBX{
C.~C.~Liu and S.~T.~Yau,
``New definition of quasilocal mass and its positivity,''
arXiv:gr-qc/0303019.
}

\lref\BSBA   {
T.~Banks and L.~Susskind,
``Brane - Antibrane Forces,''
arXiv:hep-th/9511194.
}


\lref\HsuCM{
E.~Hsu and D.~Kutasov,
``The Gravitational Sine-Gordon model,''
Nucl.\ Phys.\ B {\bf 396}, 693 (1993)
[arXiv:hep-th/9212023].
}

\lref\KutasovPF{
D.~Kutasov,
``Irreversibility of the renormalization group flow in 
two-dimensional quantum gravity,''
Mod.\ Phys.\ Lett.\ A {\bf 7}, 2943 (1992)
[arXiv:hep-th/9207064].
}

\lref\AharonyCX{
O.~Aharony, M.~Fabinger, G.~T.~Horowitz and E.~Silverstein,
``Clean time-dependent string backgrounds from bubble baths,''
JHEP {\bf 0207}, 007 (2002)
[arXiv:hep-th/0204158].
}

\baselineskip 18pt plus 2pt minus 2pt

\Title{
\vbox{\baselineskip12pt\hbox{hep-th/0405064}
\hbox{HUTP-04/A017}\hbox{MIT-CTP-3477}}
}{
\vbox{\centerline{Closed String Tachyon Condensation:}\smallskip\centerline{An Overview}}
}

\centerline{Matthew Headrick$^{a}$, Shiraz Minwalla$^b$, and Tadashi Takayanagi$^b$}

\medskip

\centerline{\sl ${^a}$Center for Theoretical Physics, MIT, Cambridge MA 02139, USA}
\centerline{\sl $^{b}$Jefferson Physical Laboratory, Harvard University,
Cambridge MA 02138, USA}

\vskip .1in \centerline{\bf Abstract}

These notes are an expanded version of a review lecture on closed
string tachyon condensation at the RTN workshop in Copenhagen in September
2003. We begin with a lightning review of open string
tachyon condensation, and then proceed to review recent results on 
localized closed string tachyon condensation, focusing on two simple systems, ${\bf C}/{\bf Z}_n$ orbifolds and twisted circle compactifications. 

\noblackbox

\Date{}

\listtoc
\writetoc

\newsec{Introduction}

Consider a classical, nonrelativistic charged particle interacting with an electromagnetic field. Let it be localized near the origin of space by a spherically symmetric potential $V(r)$ that takes the form shown in fig.\ 1. A glance at this potential carries a lot of information. For instance, it is clear that the particle has at least two stable equilibria (at $r= a$ and $r=c$) and two unstable equilibria (at $r=0$ and $r=b$). The spectrum of small fluctuations about the unstable equilibrium $r=0$ is that of an inverted harmonic oscillator, i.e.\ it is tachyonic. Displacing the particle away from $r=0$ initiates a process of tachyon condensation, and the long time behaviour is rather clear. It eventually settles down to the stable equilibrium $r=a$, while the extra energy $V(0) - V(a)$ is carried away by electromagnetic radiation.

\fig{A possible energy landscape as a function of position for a particle.}{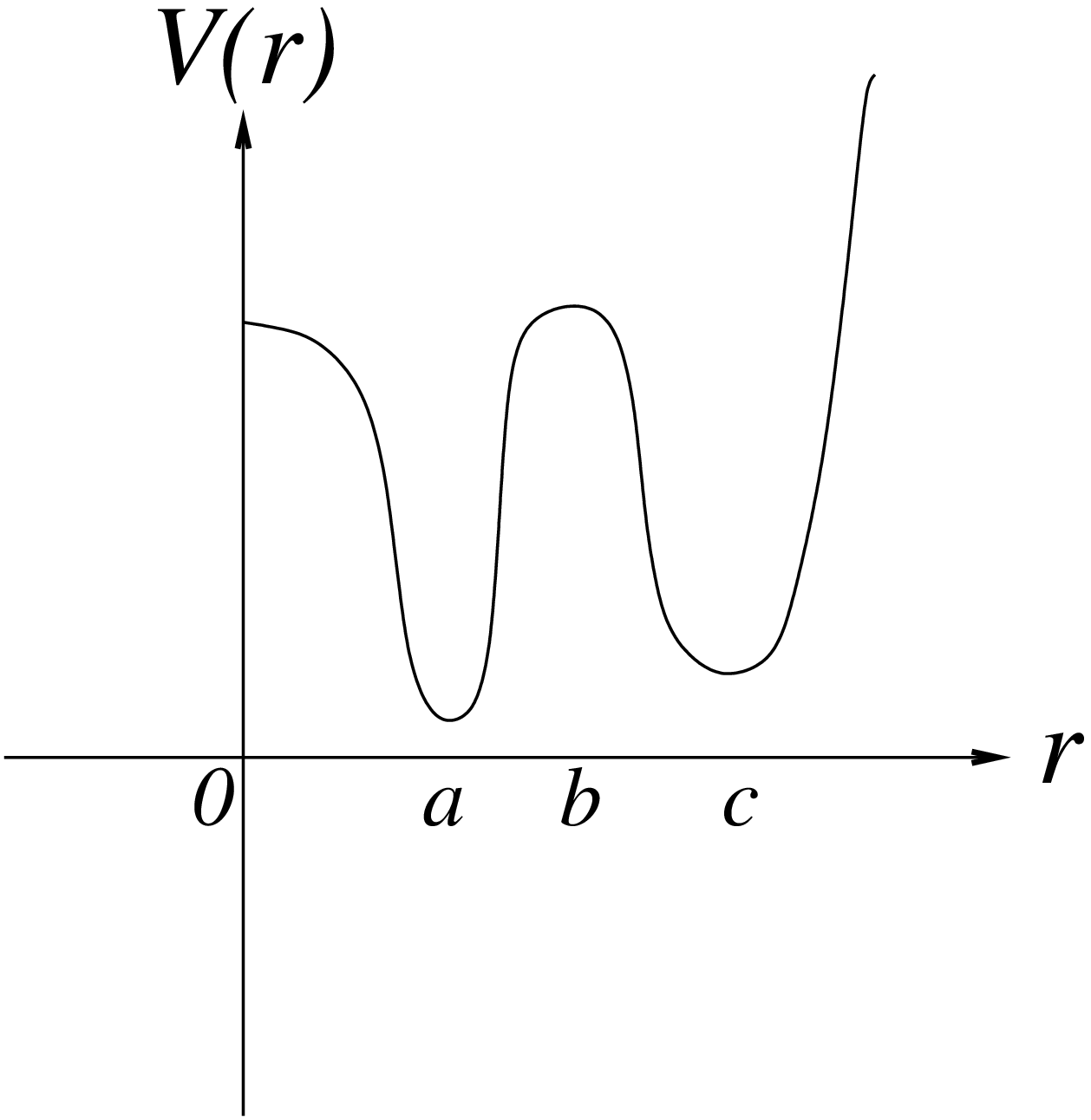}{2.5truein}

In contrast to this toy model, our understanding of the global structure of the string theory configuration space is rather limited. We do not know whether all of the different consistent string theories and their various solutions fit together as extrema of some potential in a single configuration space, like the points $r=0,a,b,c$. For example, can the bosonic string be continuously connnected to the superstrings? Recent studies of tachyon condensation in string theory may be thought of as an attempt to better understand the global structure of the string theory configuration space. These investigations study the decay of specific unstable backgrounds of string theory. Their aim is to excite the instability about the chosen unstable background (the analogue of $r=0$) and then follow the long time behaviour of the ensuing motion, to determine the nearest ``minimum" (the analogue of $r=a$).

In practice, however, exact time dependent solutions of string theory are difficult to come by. Furthermore, they sometimes carry more information than is of interest (such as the precise form of the electromagnetic radiation in our toy model). Consequently most studies of tachyon condensation circumvent the problem of solving for the exact time evolution, substituting a more tractable and possibly more interesting problem.

In our toy model, what techniques could we use to determine the endpoint of tachyon condensation about $r=0$, without solving for the full time evolution? One suggestion would be to model the backreaction of the electromagnetic radiation by a friction term in the particle evolution equation:
\eqn\damp{
{d^2 r \over d t^2}= -V'(r) -k {dr \over dt}
}
where $k$ is a positive number. A stringy equivalent of this friction term is provided by a dilaton that varies linearly in the time direction. Such a linear dilaton is present in Liouville theory. Consequently, some recent studies of tachyon condensation analyze the Liouville evolution of unstable backgrounds.

Note that in our toy model the long time behaviour of the dynamics governed by \damp\ is actually independent of $k$. We could therefore further simplify the analysis by taking $k$ to infinity, yielding the gradient flow equation
\eqn\gf{
{dr \over dt'} = - V'(r)
}
where $t= k t'$. As we argue below, the RG flow equations on the world-sheet of the string are analogous to the equations \gf. Consequently, several recent studies analyze RG flows in order to study tachyon condensation.

In this lecture we will introduce the reader to recent studies of localized closed string tachyon condensation. We will not survey all the results that have been obtained in this area, but will instead focus on two of the simplest and best studied backgrounds of type II string theory: the ${\bf C}/{\bf Z}_n$ orbifolds, and the twisted circle compactifications. In our study of these two examples we will have occasion to employ many the techniques (RG flows, Liouville flows, D-brane probes, and supergravity dynamics) that have recently been brought to bear on the problem of tachyon condensation in string theory. The reader who wishes to continue his study of tachyon condensation in other examples (including non-supersymmetric ${\bf C}^2/{\bf Z}_n$ orbifolds) is referred to the recent review by Martinec \Martinec\ (and references therein).

We begin our lecture, in section 2 below, with a lightning review of open string tachyon condensation on the world-volume of D-branes.\foot{See \refs{\TZRE,\SenNF} for comprehensive reviews of open string tachyon condensation.} Open string tachyon condensation processes share some similarities with their closed string counterparts. However, at weak string coupling 
open strings are decoupled from gravity and so open string tachyon condensation takes place on a fixed background geometry. Consequently it involves fewer conceptual complications and has been much better studied than its closed string counterpart. The review of open string tachyon condensation in section 2 will help orient us in our later study of closed string tachyon condensation. We will also see what conceptual and technical advantages localized tachyons offer over bulk tachyons (such as the type 0 or bosonic closed string tachyons). In section 3 we turn to the analysis of the simplest closed string tachyon condensation process, the decay of ${\bf C}/{\bf Z}_n$ to flat space in type II theories, and study it using a variety of methods. In section 4 we turn to tachyon condensation in a more versatile set of models, twisted circle compactifications of type II theories, which reduce in one limit to ${\bf C}/{\bf Z}_n$ orbifolds and in another to type 0 string theories. Our analysis of these models will employ Liouville flow. We also briefly study the evolution of D-branes during the tachyon condensation process. We end the lecture in section 5 with a discussion of future directions.

\newsec{Open String Tachyon Condensation
}

\subsec{Sen conjectures}

Consider a D$p$-brane in bosonic string theory. The spectrum of fluctuations of the D-brane is obtained by quantizing open strings, and includes a tachyon, implying the instability of the brane in question. It is natural to ask what the endpoint of this instability might be. Addressing this issue, Sen conjectured that {\it (i) the endpoint of homogeneous tachyon condensation on the world-volume of the D-brane is the closed string vacuum,} and {\it (ii) the condensation of inhomogeneous modes of the tachyon field leads to lower dimensional D-branes}\refs{\SenRe,\SenBA,\SenBO}.

Sen's conjectures have straightforward generalizations to the
superstring theories. The most familiar D-branes in
those theories are supersymmetric, hence stable and without tachyons on their 
world-volumes. However, open string tachyons appear on the
world-volumes of brane-antibrane systems \refs{\SenBA,\BSBA} 
and the less familiar non-BPS D-branes
\refs{\BG,\SenNB,\LR}. Sen's
conjectures concern the endpoint of tachyon condensation on these 
nonsupersymmetric brane systems.

We first consider the non-BPS D$p$-branes in the type II theories;
these exist with the ``wrong" dimensionality, i.e.\ $p$ odd in IIA and even in IIB.
Their properties are similar to those of D-branes in bosonic string theory;
in particular they are uncharged and unstable, and possess a real scalar
tachyonic field $T$ with $m^2=-1/(2\al)$. However, in
contrast to the bosonic D-branes, the tachyon effective action on unstable
type II branes is invariant under $T \to -T$, and is believed to be bounded from below. Sen conjectured that, as for bosonic branes,
homogeneous tachyon condensation on the world-volume of an unstable type II brane
ends simply in the closed string vacuum
$T=\pm T_{\rm min}$.

We now turn to inhomogeneous tachyon condensation in the same system. 
As we remarked above, the tachyon potential is invariant under 
$T \r -T$; since it also admits a minimum at a non-zero value of $T$, it is has the form of a double well. Consequently,
the tachyon field admits kink solutions that asymptote
to $-T_{\rm min}$ on the left and $+ T_{\rm min}$ on the right. 
Sen conjectured that such a kink is in fact a supersymmetric 
D$(p-1)$-brane, which may therefore be reached as the end product of 
condensation of a particular inhomogeneous mode of the tachyon.

A brane-antibrane system consists of a usual BPS D$p$-brane and
its anti-brane (i.e.\ D-brane with the opposite R-R charge).
The open strings connecting two branes (or two antibranes) are subject to
the usual GSO projections: the tachyon is projected out and the
gauge field is projected in. On the other hand, the open
strings between a brane and an antibrane are subject to the opposite
GSO projection, so the tachyon field $T$ is projected in and the gauge field
is projected out. The retained tachyon is a complex scalar field, charged oppositely under the brane and antibrane's gauge fields $A^{(+)}$ and $A^{(-)}$.
Sen conjectured that this field has a Mexican hat potential, and that homogeneous condensation to the ground state $|T|=T_{\rm min}$ results in the closed string vacuum,
while a vortex solution with winding number $n$,
\eqn\vortex{
T(r\to\infty,\theta)\sim T_{\rm min} e^{in\theta}, \qquad
\int (F^{(+)}-F^{(-)})=2\pi n,
}
is simply a supersymmetric D$(p-2)$-brane, which may therefore be obtained
as the endpoint of inhomogeneous tachyon condensation with the appropriate
boundary conditions. Finally, he proposed that a kink solution, in which the tachyon field stays real but passes over the hump at $T=0$,
\eqn\kind{
T(x\to\pm\infty)\to\pm T_{\rm min},
}
which is clearly an unstable configuration, represents an unstable D$(p-1)$-brane.

Sen's conjectures have been verified in a number of different ways, 
as we will review in the rest of this section.

\subsec{The potential: cubic string field theory}

In general, if a theory admits a potential energy landscape (as our toy example of section 1 did), then inspection of this potential is
a conceptually simple way to determine the endpoint of tachyon
condensation. Unfortunately, string world-sheet techniques are best suited
to addressing on-shell questions; string S-matrices encode information
about off-shell quantities like potentials only indirectly. Witten's
cubic open string field theory \WiSCFT, on the other hand, is an off-shell description of open string theory, and provides a potential on the open string configuration space. If we can calculate the potential of this string field theory, then inspection of it should
be sufficient to determine the endpoint of open string tachyon
condensation. This is indeed the case, as we review in this subsection.
This approach to open string tachyon condensation was poineered in
\refs{\KosteleckyNT,\SenUN,\SenCS}; for more complete references see the
reviews \refs{\TZRE,\Oh,\DeS,\ABG,\TaRE,\BonoraXP}.

The space of classical configurations of cubic open string field theory
is simply the quantum Hilbert space of the first-quantized theory.
The string field theory action associates a number to
every state in the theory of 26 free bosons plus the $b$ and $c$ ghosts on
the strip. Specifically, the action is given by \WiSCFT
\eqn\cubic{S=-\f{1}{g_{\rm o}^2} \left(\f{1}{2}\la
\Phi | Q_{\rm B} \Phi \lb+\f{1}{3}\la \Phi|\Phi* \Phi\lb\right),}
where $Q_B$ is the usual world-sheet BRST operator and $*$ is a multiplication operator
on world-sheet states; see the reviews referred to above for more details. \cubic\
enjoys invariance under the gauge transformations
$\Phi  \r \Phi  + Q_B \Psi+ \Phi *\Psi -\Psi * \Phi$ (at the linearized
level this implies a familiar statement: BRST trivial states are pure gauge).
 The equations of motion that follow from \cubic\ are
$Q_B \Phi +\Phi  * \Phi =0$ (at the linearized level, on-shell field configurations
are BRST closed).

Note that every oscillator state
in $\Phi$ is also a function of the zero-mode $x^\mu$ of every coordinate $X^\mu$ that obeys
Neumann boundary conditions. Thus each open string oscillator
state corresponds to a quantum field living on the world-volume of
the brane, and one may think of string field theory as a quantum field theory, in particular a gauge theory, with infinitely many fields propagating on the brane's world-volume.
More precisely, the string field may be expanded as
\eqn\field{
\Phi= T(x^\mu) c_1|0\lb+u(x^\mu)
 c_{-1}|0\lb+ v(x^\mu) L_{-2}c_1|0\lb+\ddd,
 }
where the lowest mode $T$ is the open string tachyon field  and
$u,v$ represent higher modes. Formally, one may
obtain an effective potential $V(T)$ for the tachyon by
integrating out all massive fields \TaE. Sen's conjecture implies
that {\it this potential takes a minimum value at one point $T=T_0$, this point represents the closed string vacuum, and its depth $V(0)-V(T_0)$ is equal to
the tension of the original D$p$-brane.}

Unfortunately, the potential in the action \cubic\ is so complicated that
it has not yet proved possible to analytically locate
$T_0$ and evaluate its energy. However this programme
has been carried through to great precision in an approximation scheme
known as level truncation \KosteleckyNT. This scheme consists of simply ignoring (i.e.\ setting
to zero rather than integrating out) all fields whose world-sheet energy
is larger than a certain fixed ``level". At the lowest non-trivial level (level 0),
the scheme retains only the tachyon, setting all other
fields to zero. At this
order the potential in \cubic\ reduces simply to
\eqn\potc{
V(T)=2\pi^2
\left(-\f{1}{2}T^2+\f{1}{3}\f{T^3}{r^3}\right),\ \ \
r=\f{4}{3\sqrt{3}},}
where we have normalized the potential energy in units of the D-brane tension.
This approximated potential has a (local) minimum at $T=r^3$;
the potential evaluated at this minimum is
$V(0)-V(T_0)=\f{\pi^2}{3}r^6\approx 0.684$ (compared with 1, as predicted by the Sen conjecture). It is more difficult to locate the
minimum and evaluate its potential at higher orders in the level truncation
scheme, and the relevant computations have been performed only numerically.
The resulting values for  $V(0)-V(T_0)$ appear to converge rapidly (but not monotonically) to that
predicted by the Sen conjecture: $0.986$
at level 4 \SenCS, $0.9991$ at level 10 \MTL, and $1.00063$ at level
28 \GRE\ (for more details see the reviews \refs{\TZRE,\Oh}). These numerical
results provide rather impressive validation of the Sen conjecture, and demonstrate that cubic string field theory, which was designed to
reproduce perturbative string amplitudes, also contains nonperturbative
information about string theory. They
also testify to the validity of the level truncation approximation; even though it is a natural
scheme for neglecting presumably irrelevant highly massive modes, it is important to keep in mind that there is no proof that it should converge to the correct answer.

In a similar fashion it has proved possible to describe an inhomogeneous
solution, depending on a single direction on the brane world-volume,
within the level truncation approximation. According to the Sen conjecture
such a lump represents a D$(p-1)$-brane, and the numerical computation of
the energies of these lump solutions again agrees rather well with this conjecture
(see \refs{\HK,\MJMT,\MSZ}; for more complete references see the review \Oh).

The Sen conjectures regarding tachyon condensation on the world-volume of
unstable D-branes in type II theory have also been
successfully been tested \refs{\BeT,\SenBK,\DeSmetJE,\IqbalST} within the level truncation
approximation scheme, using Berkovits' supersymmetric cubic string field theory \refs{\BeSFT,\BeRe}. The situation with the superstring is
qualitatively similar to that with the bosonic string; we refer the
reader to the reviews mentioned before for more details.

\subsec{RG flows and boundary string field theory}

We now turn to a rather different technique for studying tachyon
condensation, involving the analysis of RG flows on the world-sheet of the
string. We will begin this subsection with a
review of some features of world-sheet RG flows for closed as well as
open strings, and their relevance to tachyon condensation. Readers who are interested in details and complete references should refer to \HKM.

Consider a type II closed string background of the form CFT$_1+\rm{CFT}_2$,
where CFT$_2$ is any unitary conformal field theory with ${\hat c}=d$,
and CFT$_1$ is the free sigma model on $R^{9-d,1}$ with fields
$X^a$ ($a=0,\ldots,9-d$). It is not difficult to see that the
operator spectrum of CFT$_2$ determines the stability of this
background under time evolution. In fact, a conformal operator
$O$ of dimension $(\delta, \delta)$ in CFT$_2$ may be dressed by
a momentum factor from CFT$_1$ to yield a marginal operator
in the full CFT: ${\hat O}=e^{iP\cdot X}O$ where $\apm P^2/2+
\delta=1$. Thus the world-sheet operator ${\hat O}$ corresponds
to a spacetime fluctuation with squared mass $M^2=2(\delta-1)
/\apm$ and so represents a tachyon or instability when $O$ is
a relevant operator ($\delta<1$). We conclude that the existence of
relevant operators in the spectrum of CFT$_2$ implies
an instability of the corresponding string background.

The discussion of the previous paragraph is easily generalized
to open strings and boundary RG flows. A boundary operator of
dimension $\delta$ in CFT$_2$ corresponds to a spacetime particle of
squared mass $(\delta - \half)/\apm$; once again the existence of a
relevant boundary operator implies the instability of the spacetime
theory.

It is tempting to go beyond such an ``infinitesimal"
statement and to conjecture a relation between the full dynamical
evolution in string theory and renormalization group flows
on the world-sheet. Clearly, the two sides have many features
in common. A world-sheet RG flow away from an unstable string background
ends at an infrared conformal field theory that may generically
be expected to be stable. Similarly, after the dust has
settled, the dynamical process of tachyon condensation is
generically expected to decay into a stable solution of string
theory. We will now argue that, in fact, boundary RG flow equations
are a rather precise analogue of gradient flow equations for the toy model
introduced in the introduction, and so should accurately capture the late
time behaviour in the tachyon condensation process. Later in this section
we will make a similar argument for bulk RG flows.

In order to draw the analogy between RG flows and the gradient flow equation
\gf, we wish to demonstrate that spacetime energy decreases along world-sheet
RG flows. Recall that Zamolodchikov's famous
$c$-theorem \ZamolodchikovGT\ for bulk RG flows
asserts that the  $C$ function, an off-shell generalization of the
central charge, is a non-increasing function of
RG scale. A similar theorem is believed to be true for boundary RG flows.
Consider a boundary conformal field theory perturbed by
$\lambda_i O_i$, where $O_i$ is a basis of boundary operators. Within
boundary perturbation theory it has been shown that there exists a function
(sometimes called the boundary entropy) $g(\lambda)$ that \AL\
(see also \refs{\KMMB,\KO}) (1) decreases along renormalization group flows, and
(2) is proportional to the disk partition function at fixed points.
Now the disk partition function has a simple spacetime interpretation:
it is minus the spacetime energy of the open string sector
of the theory, as may be verified by computing the one point
function of the dilaton on the disk \HKMS. So we see that
(up to a proportionality constant) the $g$-function is an off-shell
generalization of spacetime energy, and that
spacetime energy decreases along boundary world-sheet flows.
Consequently RG flows are analogous to \gf\ and may be used to
study the tachyon condensation process.

We will now review the RG flow that describes the decay of a
D1-brane into a D0-brane on the circle in bosonic string theory. A D1-brane
on a circle of radius $R$ has energy $R/\apm g_{\rm s}$. On the other hand the energy
of a D0-brane on the circle is given by $1/\sqrt{\apm}g_s$. Notice
the energy of a single D$1$-brane is larger than the energy of an array
of $n$ D0-branes, provided $R ^2 > n^2 \apm$. Now consider the
theory on the world-volume of the wrapped D1-brane. The boundary operator
\eqn\tacs{
T(X)=\lambda \cos\left(\f{nX}{R}\right),
}
where $X$ is the compact direction, is relevant provided
$\alpha'(n/R)^2<1$, i.e.\
if and only if the wrapped D1-brane is more massive than an array of
$n$ D0-branes. This leads us to conjecture that the RG flow induced
by this operator (when it is relevant) ends up in an array of $n$ D0-branes.
This conjecture is rather intuitive; if the operator in \tacs\ is revelant,
$\lambda$ increases as RG flow proceeds, increasingly localizing open
strings  to the $n$ minima of the potential. It is certainly plausible that
as $\lambda \to \infty$ the endpoint of the RG flow is simply an array of
$n$ D0-branes located at these $n$ minima. Indeed this conjecture can be
rigorously verified, since the perturbation \tacs\ about the D1 background
happens to be integrable \HKM. Consequently, it is possible to follow this RG flow exactly,
and in particular to compute the boundary entropy as a function of the flow.
In particular the ratio of $g$ at the UV fixed point to $g$ at the IR fixed
point is given by 
\eqn\gfhi{
{g_{\rm UV} \over g_{\rm IR}}= {R \over n \sqrt{\apm}} =
{E_{\rm D1} \over n E_{\rm D0}},
}
in agreement with the conjectured endpoint of this RG flow.

This analysis may partially be extended to the superstring. 
Consider a D1-brane and anti-D1-brane
wrapping a circle of radius $R$. Turn on a Wilson line $\int dx^1A_1= \pi$ on the
brane, setting the gauge potential on the antibrane to zero.
As we described above, the spectrum of brane-antibrane strings includes
a complex tachyon. Since the tachyon field is charged under the relative
gauge field, this Wilson line requires $T$ to be expanded in half-integer modes on the circle.
The lightest mode of the tachyon has momentum $1/2 R$; its effective
mass is given by
\eqn\blah{
m^2 = {1 \over 4 R^2}-{1 \over 2 \apm}.
}
This is negative, and so the corresponding boundary operator is relevant, provided
\eqn\blahblah{
{2 R \over \apm g_{\rm s}} > {\sqrt2\over\sqrt{\apm} g_{\rm s}}
}
i.e.\ as long as the energy of the brane-antibrane pair is larger than the energy of an unstable D0-brane. We are thus led to conjecture that the endpoint of the corresponding
renormalization group flow is a single unstable zero brane sitting on the
circle. Note that a tachyon with half a unit of momentum around the circle
is basically a kink, so this result is consistent with the Sen conjecture.

It has, unfortunately, not yet proved possible to rigorously verify this 
conjecture. The D1-brane boundary conformal field theory
perturbed by the tachyon boundary operator is given by
\eqn\biu{
S_{\rm b} = 
\int d\tau d\theta \left({\bar \Gamma}D_{\theta} \Gamma +T(X)\Gamma\right)+ {\rm c.c.},
}
where $\Gamma=\eta+F\theta$ represents a (complex) boundary fermionic
superfield \refs{\HKM,\BSFTBA}. The fields $\eta$ and $\bar{\eta}$ are called
boundary fermions, and represent the $2\times 2$ Chan-Paton matrix upon quantization \BSFTBA. Note that this term is consistent with the fact that
the tachyon field belongs to the GSO-odd sector of the open string.
The bosonic field $F$ is an auxiliary field. In the particular case of
interest,
\eqn\Tf{
T(X)= \lambda \cos {X \over 2 R}.
}
Rigorous verification of the Sen conjecture in this case would require 
some exact results on the IR behaviour (for instance the IR value of 
the $g$ function) of \biu, \Tf.

As we described above, the 
$g$-function in the middle of RG flow can be considered as an off-shell energy
in open string theory. This statement might lead us to search for 
an off-shell formulation of string field theory, whose action is some
generalization of the $g$ function. Although we will touch on this 
topic only very briefly, indeed such a formulation has been found, and
is known as boundary string field theory, or BSFT 
\refs{\KMMB,\GeSh,\KMMS,\TsB}
(originally called background independent string field theory 
\refs{\WiBSFT,\Sh}; see the reviews \refs{\Oh,\Ue}\ for more details). 
One convenient definition\foot{This is
equivalent to the original definition of the background
independent string field theory \WiBSFT\ via a redefinition of
parameters $\lambda_i$.} \KMMB\ of the BSFT action $S$ is given by
\eqn\deb{
\f{\de S}{\de \lambda^j}=\beta^i G_{ij}.
}
The parameters $\lambda^j$ are the coefficients of possible boundary perturbations as
before\foot{In actual computations we cannot consider irrelevant
perturbations because they are not renormalizable. Even though
this is the most serious unsolved problem in BSFT, we can usually
get the correct results on tachyon condensation exactly without
taking them into account.} (e.g.\ $T(X)$ in \tacs), whose beta
functions are given by $\beta^j$, and $G_{ij}$ is a boundary
theory analogue of Zamolodchikov's metric. $S$ may be argued to be 
a generalization of the $g$ function; in particular it may be shown 
that $S$ evaluates on-shell to the disk world-sheet partition function.
Indeed, in the case of the superstring it turns out that this statement
is true even off-shell: a simple solution to \deb\ is obtained by 
setting $S=Z$ where $Z$ is the world-sheet disk partition function \refs{\BSFTBA,\KMMS,\TsB,\BSFTBV}.

Boundary string field theory has been employed to good effect in
the study of open string tachyon condensation. For instance it has been 
shown that the profile of the tachyon potential on an
unstable D-brane in the superstring is given by BSFT as $V(T)\propto e^{-T^2/4}$. The
closed string vacuum corresponds to $|T|=\infty$ (note that this $T$ is related by a non-linear field redefinition to the field that appears in the tachyon potential of cubic string field theory). 

Finally, let us point out that inverting the relation \deb,
and applying the definition of the beta function, yields the gradient 
flow equation 
\eqn\rgf{
\f{\de\lambda^i}{\de(-\ln\Lambda)} = -G^{ij}\f{\de S}{\de\lambda^j},
}
where $\Lambda$ is the renormalization scale.
Hence the analogy between RG flow and \gf\ is clearest in the 
language of boundary string field theory.

\subsec{Time evolution: rolling tachyons}

Quite remarkably, it is possible to solve exactly for the 
classical time evolution of a decaying D-brane. The solution is called the rolling tachyon 
\refs{\SenRO,\SenTM} or S-brane \refs{\GSS,\MSY}.
The homogeneous decay can be simply described by perturbing the D-brane conformal field theory by the boundary operator \refs{\SOP,\LNT,\GSL}\foot{Following the general convention, we have set $\apm=1$ in this subsection.}
\eqn\wilsont{
S_{\rm b}=\lambda\
 \int_{\de\Sigma} d\tau e^{X^0}.
}
(More precisely, this solution is known as a half S-brane.)
To understand why this is a conformal field 
theory, we first note that \wilsont\ may be obtainted from the BCFT \SenRO 
\eqn\wilsontt{
S_{\rm b}=\lambda'\ \int_{\de\Sigma} d\tau \cosh(X^0(\tau))
}
(known as a full S-brane) by taking the limit $ \lambda' \to 0$, $X^0 \to \infty$ with $\lambda' e^{X^0}$ held fixed. Now upon analytically continuing $X=iX^0$, \wilsontt\ turns 
into the marginal boundary perturbation \tacs. This perturbation is in fact
exactly marginal \refs{\SenBO,\CKLM,\RSM}, as it can be rotated into a Wilson line
by using the exact level 1 $SU(2)$ current algebra present in the open 
string theory at $R=n=1$. We conclude that the deformation 
\wilsont\ is exactly marginal, and represents a boundary 
conformal field theory, i.e.\ an exact time dependent solution of
classical open string theory.  The physical interpretation of this 
solution is clear: at large negative times \wilsont\ represents the 
D-brane perturbed very slightly by a tachyon operator. This operator 
then proceeds to grow in time according to its equation of motion. Thus 
this solution corresponds to tachyon condensation.

Now that we have the exact solution to the time decay of the tachyon, it might 
seem that we should easily be able to determine its long time behaviour. 
In fact this issue turns out to be rather subtle. \wilsont\ represents a 
boundary conformal field theory, and so a theory of open strings. How is this
consistent with the Sen conjecture, which holds that the final state can contain no open strings? An important clue is obtained by computing the energy emitted in closed string 
radiation in the background \wilsont. The presence of a D-brane with 
rolling tachyons leads to a linear source, a tadpole, for each closed string 
mode \refs{\OS,\CLL}, and the amount of radiation can be computed by 
the one-point function of the closed string vertex operator on the disk. The amplitude to emit a 
state of energy $E$ is proportional to \LLM
\eqn\onep{
\la e^{iEX^0} \lb_{\rm disk}= e^{-iE \log \lambda } \f{\pi}{\sinh(\pi E).}
}
In the high energy region, the square  of the amplitude \onep\ decays exponentially like
$e^{-2\pi E}$. Consequently the D-brane decay process populates the 
closed string states in a thermal fashion, at an effective temperature
that is twice the Hagedorn temperature. However, not all closed string 
states are thus excited; the D-brane couples only to left-right symmetric
states, and the density of excited states scales like 
$\rho(E)\sim e^{2\pi E}$, the square root
of Hagedorn growth. Note that the exponential growth in the density  
of states exactly cancels the exponential supression in the squared 
amplitude. Consequently the amount of energy emitted into closed strings, at 
tree level, is controlled by subleading powers of $E$ in the density of 
states. Detailed calculations demonstrate that the energy of emitted 
radiation from unstable D$p$-brane scales like 
$\int dE\,E^{-p/2}$, and therefore diverges for 
$p=0,1,2$. The energy radiated is formally finite for $p>2$ \LLM; however 
this finiteness is only power-law and probably disappears at higher loop.

Let's focus on the case $p=0$. The computation reviewed
above strongly suggests that the D0-brane dumps all of its energy into 
closed string radiation in a time of order the string scale. The divergence in the
computation appears simply to reflect the fact that the energy of a D0-brane 
scales like $1/g_{\rm s}$ and so diverges in perturbation theory. 
However the divergence also formally reflects a breakdown in perturbation 
theory, so it is unclear if more can be said about the process. 

Does the uncontrolled production of closed strings ensure that the boundary 
conformal field theory \wilsont\ breaks down at a time of order string 
scale? Naively that would appear to be the case. However Sen has recently 
conjectured instead that the production of closed strings
has a dual description in terms of open strings 
\refs{\SenOC,\SenDU,\SenZM} (see also related arguments \refs{\MSY,\GRB}). 
This conjecture appears to be supported by 
computations in two-dimensional string theory. However, aspects 
of it are controversial, and the last word on the subject is perhaps yet to be said. 

Several other issues surrounding this time evolution process remain unclear. 
These include questions about the endpoint of the process 
with high dimensional branes, and the interpretation of similar computations 
in linear dilaton backgrounds \KLMS. We will not attempt a discussion of these 
cutting-edge issues here.

\subsec{Open versus closed string tachyon condensation}

In the last four subsections we saw that the problem of open string tachyon condensation on D-branes has been successfully addressed using several different techniques. Let us now recall the various approaches used, and ask if they generalize to closed string theories.

The direct study of the dynamical decay of closed string tachyons depends on finding exact solutions, which is usually difficult. This approach may also run into conceptual hurdles. As we described above, it proved rather difficult to extract the qualitative properties of the long time behaviour of open string tachyon condensation from the exact solution, even though we had a good guess for the answer. This process may be even more difficult in the case of closed strings. Nonetheless exact time dependent solutions, when available, are exciting, and deserve to be studied in detail \refs{\StromingerFN,\DaCunhaFM}. We will touch upon attempts to construct and analyze such solutions in the final section of this lecture.

In subsection 2.2 we reviewed the construction of the open string tachyon potential using string field theory. It seems likely that any attempt to effect similar constructions in closed string field theory will be considerably more difficult. The reasons for this expectation are conceptual as well as technical. At the classical level, open string field theory is ``merely" a field theory, albeit a very complicated one involving an infinite number of fields. In particular all dynamics takes place in a static background, and every configuration may unambiguously be assigned an energy. Closed string theories are always theories of gravity, and spacetime is dynamical in such theories. Consequently the notion of energy is rather subtle. In particular, several recently studied examples of closed string tachyon condensation involve processes that drastically modify the asymptotics of spacetime; in such cases it is not clear whether it is possible, even in principle, to compare the energies of the initial and final solutions. A satisfactory potential function on the space of  closed string configurations would be a very exciting object indeed, but finding it would involve overcoming several conceptual barriers.

At the technical level, the existing formulations of  closed string field theories are more complicated than open string field theories. In particular, Zwiebach's covariant closed string field theory \ZwiebachIE\ (which has so far been constructed only for the bosonic string) has non-polynomial interactions, and it is unclear if any simple approximation scheme like level truncation will apply to it. In a recent study, however, Okawa and Zwiebach \OkZw\ report exciting progress in this direction.

Of the techniques used successfully to study open string tachyon condensation, the one that generalizes most simply to closed strings is that of world-sheet RG flows. As we reviewed above, the instability of a spacetime theory implies the existence of a relevant operator, and hence an instability of RG flow, on the world-sheet of the string. The problem of following this RG flow to the IR is mathematically well posed. In the case of open strings the connection  between the RG flow and the problem of tachyon condensation was established by demonstrating that the world-sheet RG flow equations are analogous to the gradient flow equation \gf, and in particular that spacetime energy decreases along RG flows. It turns out that a similar (though weaker) result is true of bulk RG flows seeded by localized closed string tachyon vertex operators. In particular, one may show \GuHeMiSc\ that whenever such an RG flow connects two spaces whose asymptotics are similar enough that their energies can be compared, the ADM energy of the IR solution is lower than that of the UV solution. Thus spacetime energy decreases along bulk world-sheet RG flows, for all flows for which this statement may be sensibly formulated. Consequently it appears that, like their boundary counterparts,  bulk RG flows may also be thought of as analogues of the gradient flows \gf.

As we will review in these lectures, RG flows have been widely used to study closed string tachyon condensation. Another technique that has been used is the so-called Liouville flow \KutasovPF, which describes a continuous path (presumably from higher to lower ``energy") through the space of string configurations. 

In this review, following the example of most recent studies of closed string tachyon condensation, we will focus on systems in which the tachyons are localized on a proper submanifold of the spacetime, for example an orbifold fixed plane. Such localized tachyons offer several advantages over bulk tachyons like those of the bosonic and type 0 theories. First, since open string tachyon condensation leads to the decay of the D-branes on which the open strings live, by analogy it is may turn out that bulk closed string tachyons lead to the decay of the spacetime itself, and with it the disappearance of all closed string states. (A non-perturbative analogue of this is Witten's ``bubble of nothing" decay of non-supersymmetric Kaluza-Klein compactifications \WiBON.) Using conventional tools it would seem difficult to study such a drastic transition, whose endpoint is not even a string theory. In the case of localized tachyons, whatever happens when they condense should initially be confined to the region in which they are localized. This gives the researcher a better chance of following the process.

Second, the problem of tachyon condensation, which is a dynamical process, is often loosely equated with the problem of finding the ground state (or ``true vacuum") of the system. Such an equation, while quite natural, presumes some mechanism for dissipating the energy released in the condensation. For example, the particle in the toy example of section 1 was able to find its ground state by dissipating its extra energy to infinity in the form of radiation. Similarly, a system with tachyons localized in a non-compact space will be able to dump the energy released by their condensation into the surrounding space (presumably in the form of closed string radiation), and thereby relax to its ground state. In contrast, in the case of bulk closed string tachyons, conservation of energy severely complicates the issue of whether condensation leads to the true vacuum of the theory, if it has one.

Finally, while world-sheet RG flow has proven to be such a powerful tool for studying both open and localized closed string tachyon condensation, it is unclear whether it has any simple relation to the dynamical process of tachyon condensation in the case of bulk closed string tachyons. The reason is the Zamolodchikov $c$-theorem \ZamolodchikovGT, which leads us to expect the central charge of the world-sheet theory to decrease under any flow that is homogeneous in the target space. On the other hand, in any spacetime process like tachyon condensation it is pegged at $c=15$ or 26; after all, a spacetime process from the point of view of the world-sheet is just another $c=15$ or 26 conformal field theory.

In the last section of this review we will return to a brief discussion of the problem of bulk tachyon condensation, including a possible way around the $c$-theorem.

\newsec{Decay of ${\bf C}/{\bf Z}_n$}

\subsec{Introduction}

The ${\bf C}/{\bf Z}_n$ orbifold of type II string theory, which we will describe in this subsection, is an ideal laboratory for studying closed string tachyon condensation for several reasons. Its tachyons are localized on a co-dimension 2 surface, the orbifold fixed plane, far from which supersymmetry is restored. Furthermore, it is connected by smooth deformations to a supersymmetric vacuum, namely type II string theory in flat space. Finally, and perhaps most importantly, it is extremely simple. Because of these features, ${\bf C}/{\bf Z}_n$ has become a paradigm for the study of closed string tachyons ever since the seminal paper \AdPoSi.

The orbifold \refs{\DabholkarAI,\LoweAH} is a $\hat c=2$ SCFT constructed by twisting the free theory, with its flat target space $\bf C$, by a ${\bf Z}_n$ subgroup of its $U(1)$ rotational symmetry. Due to the existence of spacetime fermions, this $U(1)$ is a double cover of the geometrical $SO(2)$ rotation group: the rotation angle has periodicity $4\pi$. Therefore the generator of the ${\bf Z}_n$ subgroup is $e^{4\pi iJ/n}$, where $J$ is the angular momentum in this plane. If $n$ is even then this subgroup includes $e^{2\pi iJ}=(-1)^F$, where $F$ is the (spacetime) fermion number, and among the twisted sectors are type 0 fields, including a bulk tachyon. To avoid a discussion of bulk tachyons, we will take $n$ odd. In this case it may be more useful to think of the element
\eqn\orbgen{
(-1)^Fe^{2\pi iJ/n}
}
(the square root of $e^{4\pi iJ/n}$) as the group's generator.

The target space of the ${\bf C}/{\bf Z}_n$ orbifold is a cone, rather like the space transverse to a cosmic string. ADM mass for co-dimension 2 objects is measured by the deficit angle at infinity; in this case the mass (or, more precisely, tension, since the orbifold is extended in seven dimensions) is (see e.g.\ \refs{\GuHeMiSc,\Dabh})
\eqn\energyden{
M_{{\bf C}/{\bf Z}_n} = {2\pi\over \kappa^2}\left(1-{1\over n}\right).
}
Just as a cosmic string is a soliton of some matter-gravity equations of motion, the orbifold should be thought of a stringy soliton, a static solution to the full string equations of motion with a localized concentration of energy at the fixed point. The dependence of the mass \energyden\ on the gravitational coupling shows that the orbifold, like for example the NS5-brane, is a closed string soliton. Unlike the NS5-brane, however, it is non-BPS: no spacetime spinors are invariant under the rotation, so the spacetime supersymmetry is completely broken.

The string spectrum can be derived straightforwardly in either the RNS or the Green-Schwarz formalism. Here we will sketch the derivation of the most important features using the RNS formalism, leaving the details to appendix A. Let us assume that it is the $X^8$-$X^9$ plane that is being orbifolded; for simplicity we will take the other eight world-sheet fields $X^i$ ($i=0,\dots,7$) to be free, so that the spacetime is $M^{7,1}\times{\bf C}/{\bf Z}_n$. It is convenient to pair $X^8$ and $X^9$ and their superpartners into complex combinations:
\eqn\fpair{
Z = {1\over\sqrt2}\left(X^8 + iX^9\right), \qquad
\psi = {1\over\sqrt2}\left(\psi^8 + i\psi^9\right), \qquad
\tilde\psi = {1\over\sqrt2}\left(\tilde\psi^8 + i\tilde\psi^9\right).
}
In the $k$th twisted sector, the periodicities of these fields are twisted, while the rest have their usual periodicities:
\eqn\tbc{\eqalign{
Z(\w+2\pi) &= e^{-2\pi ik/n}Z(w), \qquad\,\,\, X^i(w+2\pi) =  X^i(w) \cr
\psi(w+2\pi) &= \pm e^{-2\pi ik/n}\psi(w), \qquad \psi^i(w+2\pi) = \pm \psi^i(w), \cr
\tilde\psi(w+2\pi) &= \pm e^{-2\pi ik/n}\tilde\psi(w), \qquad \tilde\psi^i(w+2\pi) = \pm \tilde\psi^i(w).
}}
The upper and lower signs refer respectively to R and NS periodicities. As usual the left-moving fermions must be either all R or all NS in order for the left-moving world-sheet supercurrent $T_F=i\sqrt{2/\alpha'}\psi^\mu\partial X_\mu$ to be well defined; similarly for the right-movers.

According to \tbc, the twisted strings must stretch in order to move away from the fixed point, so they do not have zero modes for moving in the orbifold directions. Thus the twisted string fields live ``on" the co-dimension two fixed plane, in the same sense that open string fields live on the D-brane world-volume. Since the spectrum of untwisted strings is supersymmetric, the supersymmetry breaking is localized near the fixed point, and away from it the theory is locally type II strings in flat space.

The twist number $k$ is valued in ${\bf Z}_n$ (i.e.\ $k\sim k+n$), and it is conserved (mod $n$) in interactions. It is therefore possible (and useful) to construct a symmetry for which it is the charge; this is the so-called ``quantum" ${\bf Z}'_n$ symmetry. Under an element $k'\in{\bf Z}'_n$, the wave function of a string in twisted sector $k\in{\bf Z}_n$ is multiplied by $e^{2\pi ikk'/n}$. Of course, the original ${\bf Z}_n$ symmetry by which we orbifolded is gone, since only states that are neutral under it survive the orbifold projection.

The boundary conditions \tbc\ shift the modings of the $Z$, $\psi$, and $\tilde\psi$ oscillators; in particular, the $\alpha$ and $\psi$ mode numbers are shifted by $-k/n$, while the $\tilde\alpha$ and $\tilde\psi$ mode numbers are shifted by $k/n$. A straightforward computation gives the total zero-point energy in the NS case as $-{1\over2}(1-|k|/n)$, where in this paragraph we choose the twist number $k$ to lie in the range $-n/2<k<n/2$. On the other hand, world-sheet supersymmetry is unbroken by the R boundary conditions for all $k$, guaranteeing the vanishing of the zero-point energy. Any tachyons must therefore be NS-NS states.

Usually the type II GSO projection projects out the NS ground state, but here the following useful rule of thumb comes into play: {\sl When twisting by a symmetry involving $(-1)^F$, the GSO projection gets reversed in the odd twisted sectors.} (Recall for example the ${\bf Z}_2$ orbifold of the type II string by $(-1)^F$ itself, which produces the type 0 string, with the (NS$-$,NS$-$) and (R$-$,R$\pm$) states arising in the twisted sector.) According to this rule, which will be rigorously justified for the case of ${\bf C}/{\bf Z}_n$ in appendix A, the ground state is projected {\sl out} for even $k$ but {\sl in} for odd $k$. For even $k$ we must act on it with a fermionic raising operator such as $\psi_{-1/2-k/n}\bar{\tilde\psi}_{-1/2-k/n}$ or $\bar\psi_{-1/2+k/n}\tilde\psi_{-1/2+k/n}$. The lowest projected-in state has mass given by
\eqn\tspectrum{
m_k^2 = -{2\over\alpha'}
\cases{\left(1-\f{|k|}{n}\right), & $k$ odd \cr\f{|k|}{n}, & $k$ even}, \qquad
(|k|<n/2).
}
(This is the mass with respect to the remaining $7+1$ dimensional Poincar\'e symmetry of the orbifold.) We see that every twisted sector contains at least one tachyon. Those with $k$ odd and $|k|/n<1/3$ also contain other tachyons, obtained by acting on the lowest one with the bosonic raising operator $\alpha^Z_{-k/n}\bar{\tilde\alpha}^Z_{-k/n}$ ($k>0$) or $\bar\alpha^Z_{k/n}\tilde\alpha^Z_{k/n}$ ($k<0$).

We will refer to the vertex operator for the lowest tachyon with twist $k$ as $V_k$ (we will not need the precise expression here; it may be found for example in \DaIqRa), and the corresponding spacetime field as $T_k$. Both are complex (recall that they are charged under the quantum symmetry), and their complex conjugates are
\eqn\conjugates{
V_k^* = V_{-k}, \qquad T_k^* = T_{-k}.
}

The fundamental question we would like to answer is, What happens when these tachyons condense? Because they are localized near the orbifold fixed point, when they condense the effect should initially only be felt in that region. There are various scenarios one could consider: a throat or a baby universe could be created, or a tear or bubble of nothing could open up. A less dramatic possibility is that the topology of the spacetime does not change, but instead its geometry simply smooths out, with the curvature singularity at the fixed point replaced by a smooth cap. If this cap then expands into the surrounding space, diluting the curvature and eventually growing to infinite size, then we would say that the endpoint of the tachyon condensation is flat space (although, strictly speaking, due to the non-compactness of the space that ``endpoint" would not be reached in any finite time). This last scenario is closely analogous to what happens when tachyons on D-branes condense: at the endpoint supersymmetry is restored, and those open strings ending on the decaying branes (the analogues of the twisted strings) are lifted out of the perturbative spectrum, while any bulk open strings (the analogues of the untwisted strings) persist.

This scenario was first conjectured to be the correct one by the authors of \AdPoSi, and is therefore referred to as the ``APS conjecture". They gave substantial evidence in its favor, and quite a bit more has accumulated since then. In this section we will give a brief overview of some of this evidence. The strongest piece is an exact construction of the RG trajectory of the world-sheet theory perturbed by the most relevant tachyon vertex operators, the ones with $k=\pm1$. As shown in \Va, this flow has as its IR fixed point the sigma model onto the plane. This construction is described in subsection 3.2. An exact solution for the flow can also be obtained in the sigma model limit, which gives intuition about how the flow proceeds at late stages when the geometry becomes smooth and the curvatures are small in string units \GuHeMiSc. This solution is described in subsection 3.3.

As explained in section 2, however,  in the closed string context the relation between world-sheet RG flow and the actual dynamical tachyon condensation process is conjectural, even more so than in the open string context, so the above results on RG flow cannot be taken as sure indicators of the endpoint of the condensation. Unfortunately there are fewer results about the time evolution of the system, which is likely to be quite complicated. Although the technology does not yet exist to follow the dynamics on time scales comparable to the string length, it is possible to gain some support for the conjecture by looking at its behavior on much shorter and much longer time scales. When the tachyon has a very small expectation value, we can see how the geometry is deformed as seen by a D-brane probe; the result \refs{\AdPoSi,\chicago} is that the tip of the cone is indeed smoothed out, and the size of the cap grows with the tachyon expectation value. This analysis is described in subsection 3.4. Furthermore, if a tear or other singularity does not form in the space, then the curvature scales will eventually become large enough for supergravity to be valid. At that point, as explained in subsection 3.5, we can confidently predict the future evolution of the system, and show that a bubble of flat space expanding at the speed of light is its inevitable late time behavior \refs{\GrHa,\He}.

Although we will not review it here, there has also been progress towards finding an effective potential for the tachyons. One such potential was proposed in \DhVa\ on the basis of the so-called $tt^*$ geometry for the chiral ring of the system. A level truncation scheme closely analogous to that described in subsection 2.2 was applied to the non-polynomial closed string field theory to find critical points of the tachyon potential for ${\bf C}/{\bf Z}_n$ orbifolds of the bosonic string \OkZw. Finally, off-shell interactions were deduced on the basis of scattering amplitudes for light tachyons in ${\bf C}/{\bf Z}_n$ at large $n$ \DaIqRa\ (see also \SarkarDC).

It should be noted that the evidence presented here is strong but not conclusive in favor of the APS conjecture, and alternative proposals do exist, although we will not have time to review them here. In particular, the authors of \chicago\ have argued based on an analysis of the theory's chiral ring that, at least under world-sheet RG flow, the spacetime does in fact undergo a topology change, becoming a disjoint union of flat space (with type II strings) and several cones (with type 0 strings).

\subsec{RG flow: GLSM analysis}

Although the ${\bf C}/{\bf Z}_n$ orbifold breaks spacetime supersymmetry, it preserves ${\cal N}=(2,2)$ supersymmetry on the world-sheet. Furthermore, the most relevant operator in each twisted sector is a chiral primary operator. We can therefore use the power of supersymmetry to study world-sheet RG flows seeded by these vertex operators. This is what Vafa did in his construction \Va\ of these RG flows using a gauged linear sigma model (GLSM), which we will review in this subsection. He also showed, using Hori-Vafa mirror symmetry \HoVa, that the process has an elegant dual description in terms of a Landau-Ginzburg theory, which we will not review here.

The coupling constant $e$ for a gauge theory in two dimensions has units of mass, so the theory is free in the ultraviolet. Typically, one uses a GLSM to construct an interesting conformal field theory as its infrared fixed point \WittenYC. In this case we will construct an entire RG flow, not just a single fixed point, and the strategy is rather clever. We will choose a GLSM whose infrared fixed point is ${\bf C}$, so we have a flow from a free gauge theory in the UV to a free sigma model in the IR. Moreover, by construction this flow will pass close to the unstable ${\bf C}/{\bf Z}_n$ fixed point, and by taking an appropriate limit of the parameters in the GSLM we can make it pass right through that fixed point. In this limit the flow actually breaks into two consecutive flows: the first from the free gauge theory to ${\bf C}/{\bf Z}_n$, and the second (which is the one of interest to us) from ${\bf C}/{\bf Z}_n$ to ${\bf C}$. Finally, we will show that in the initial stages of this second flow the deviations from the ${\bf C}/{\bf Z}_n$ CFT are precisely those we would get by perturbing it by the most relevant vertex operator, namely the chiral primary operators $V_1$ and $V_{-1}$. This proves that the infrared physics of the ${\bf C}/{\bf Z}_n$ theory perturbed by these operators is indeed the free sigma model.

The GLSM we need is very simple. It has a $U(1)$ gauge group and two chiral superfields $\Phi_1$ and $\Phi_{-n}$.\foot{To follow the argument as presented here it is not necessary to be familiar with the details of ${\cal N}=(2,2)$ supersymmetry, such as the definitions of the various types of superfields, although certain points will have to be taken on faith. The necessary background material is explained in \refs{\HoVa,\WittenYC}.} The subscripts indicate their respective charges, so that under gauge transformations they transform as
\eqn\ft{
\Ph_1 \to e^{i\a} \Ph_1,\qquad
\Ph_{-n} \r e^{-in \a} \Ph_{-n}.
}
The gauge field $v_\mu$ is the lowest component of a vector superfield $V$ (working in Wess-Zumino gauge), and the field strength $v_{01}$ resides in the super-field strength $\Sigma$, which is a twisted chiral superfield. The action is
\eqn\act{
S = {1\over 2 \pi}\int d^2\s\,\left[
\int d^4\t \left({ \bar \Ph_1} e^V \Ph_1+{ \bar \Ph_{-n}} e^{-nV} \Ph_{-n}-{1 \over 2 e^2} |\Sigma|^2\right)
-{\rm Re}\,t\!\int d^2\!\ti\theta\,\Sigma
\right].
}
The complex coupling constant $t=r+i\theta$ parameterizes both  the Fayet-Ilipoulos coupling
$r$ and the theta angle. The quantum theory generated by \act\ is super-renormalizable; in fact, thanks to the high degree of supersymmetry, all correlators are rendered finite after a simple one-loop renormalization of the FI term:
\eqn\fit{
r(\Lambda) = -(n-1) \ln\left({\Lambda \over \Lambda_0}\right),
}
where $\Lambda_0$ is defined to be the energy scale at which $r$ vanishes.

Now let us consider the low energy dynamics of \act. The terms involving the auxiliary gauge field $D$ are
\eqn\dterms{
{1 \over 2 \pi} \int d^2\s
\left( {D^2 \over 2 e^2}+D(|\ph_1|^2-n|\ph_{-n}|^2 -r) \right),
}
where $\ph_1$, $\ph_{-n}$ are the lowest components of $\Ph_1$, $\Ph_{-n}$ respectively. At energies small compared to $e$ we may restrict attention to fluctuations on the manifold of supersymmetric, zero-energy configurations satisfying
\eqn\ze{
-\f{D}{e^2}=|\ph_1|^2-n|\ph_{-n}|^2-r=0.
}
The moduli space of classical vacua is parameterized by $\ph_1$ and $\ph_{-n}$ constrained by \ze, modulo gauge transformations: the dynamics is that of a one complex dimensional supersymmetric sigma model.

According to \fit, at energies small compared to $\Lambda_0$, the FI parameter $r$ is positive and large. Since according to \ze\ $\ph_1$ must be non-zero, we can use the gauge freedom to make it real and positive, and use \ze\ to solve for it in terms of $\ph_{-n}$:
\eqn\phione{
\ph_1 = \sqrt{n|\ph_{-n}|^2 +r}.
}
We then plug this solution into the kinetic terms,
\eqn\kintsf{
{1 \over 2 \pi}
\int d^2\s ( -\CD^\m \ph_1 \CD_\m \bar\ph_1 -\CD^\m \ph_{-n} \CD_\m \bar\ph_{-n}),
}
and classically integrate out the gauge boson $v_\m$. (When $r$ is large $\ph_1$ is also large, so the gauge boson is very massive, and since the gauge theory is free in the UV the classical approximation is valid in this limit.) We find that the dynamics in the limit $r\to\infty$ is governed by the flat sigma model $S = -{1\over 2 \pi} \int d^2\!\s\,\p_\m \ph_{-n} \p^\m { \bar \ph_{-n}}$. (Appropriate GSO projections can also be imposed, although we won't describe the details here. Suffice it to say that the GLSM possesses chiral ${\bf Z}_2$ R-symmetries that reduce at the conformal fixed points to the operators, explained in appendix A, used to impose the GSO projections.)

If $\Lambda_0\ll e$ then there also exists an intermediate range of energy scales $\Lambda_0\ll\Lambda\ll e$ for which \ze\ is valid but $r$ is large and negative. In this case it is $\ph_{-n}$ that we can make real and positive and integrate out along with the gauge field. Now $\ph_1$ parametrizes the target space with a flat metric. Note, however, that our gauge choice leaves unfixed a residual ${\bf Z}_n$ group of gauge transformations, generated by
\eqn\unfixexgt{
\ph_1 \r e^{{2 \pi i/ n}} \ph_1.
}
Consequently the theory in this range of energy scales is nothing but the ${\bf C}/{\bf Z}_n$ conformal field theory. In order to make this statement exact, we need to take the limit $r\to-\infty$, while keeping the energies small compared to $e$. Thus in the limit $e/\Lambda_0\to\infty$, the RG flow separates into two stages: in the first the theory flows from a free gauge theory to ${\bf C}/{\bf Z}_n$, and in the second from ${\bf C}/{\bf Z}_n$ to ${\bf C}$.

In order to make the story complete, we need to show that this second flow is precisely the one that is seeded by a tachyon vertex operator of the ${\bf C}/{\bf Z}_n$ sigma model, and in particular by $V_1$ and $V_{-1}$. We first note that, since the flow preserves supersymmetry, it must be seeded by one of the chiral primary operators. To find out which one, let us see how the physics as described by the GLSM changes as $r$ increases from minus infinity. When it is large and negative but finite, the gauge theory admits fractional instantons \WittenYC, vortices in which the phase of $\ph_{-n}$ winds at infinity on the (Euclidean) world-sheet. In such a configuration, we cannot (as we did in the $r\to-\infty$ limit above) make $\ph_{-n}$ everywhere real and positive. The path integral includes a sum over sectors containing arbitrary numbers of instantons and anti-instantons, and integrals over their positions. Now, due to the configuration of the gauge field, the phase of $\ph_1$ jumps by $-2\pi/n$ upon circling a fractional instanton. Therefore the effect of an instanton on a correlator is reproduced in the ${\bf C}/{\bf Z}_n$ sigma model by the insertion of $V_1$, which contains a twist field that enforces precisely this behavior, i.e.\  \tbc\ (with $k=1$). The sum over instantons and anti-instantons that occurs in the GLSM path integral is reproduced by perturbing the ${\bf C}/{\bf Z}_n$ sigma model action by
\eqn\twd{
\int d^2\sigma\,e^{(r + i \theta)/n} V_1 + \rm{c.c.}
}
The coefficient of $V_1$ is the GLSM action of a single instanton.

By replacing the chiral superfield $\Phi_1$ in the GLSM by one of charge $k$, we can see what happens when the RG flow is seeded by $V_k$ and $V_{-k}$, rather than $V_1$ and $V_{-1}$. The running of $r$ \fit\ becomes
\eqn\newfit{
r(\Lambda) = -(n-k)\ln\left({\Lambda\over\Lambda_0}\right),
}
so we must require $k<n$. The $D$-term constraint \ze\ becomes
\eqn\newze{
k|\ph_k|^2-n|\ph_{-n}|^2-r=0,
}
so we must also require $k>0$. Finally, we must require $k$ to be odd in order to be able to impose a GSO projection. (Note that these restrictions imply no loss of generality on the choice of tachyon vertex operator.) Then, following the same logic as above, we find that the IR limit of the RG flow is the lower-order orbifold ${\bf C}/{\bf Z}_k$.

\subsec{RG flow: gravity regime}

The argument of the previous subsection showed that the endpoint of RG flow seeded by the tachyon vertex operator $V_1$ is flat space. However, the question of how the flow effected the transformation of the cone into the plane remained a bit obscure. At early RG ``times", the ${\bf C}/{\bf Z}_n$ CFT is perturbed by the tachyon vertex operator, which is a twist field and so non-local in spacetime, so we should not expect any geometric picture. In the late stages of the flow, in order to find out what is happening to the geometry we should carefully integrate out the gauge field at large but finite values of $r$. It turns out that the metric for $\ph_{-n}$ is actually flat only in the region $|\ph_{-n}|^2\ll r$, and remains conical for $|\ph_{-n}|^2\gg r$ (see appendix B of \MiTa\ for a detailed discussion). As $r$ grows according to \fit, this flat region grows in size, so that from the point of view of an observer at any fixed distance from the origin the space does indeed eventually become flat (although strictly speaking the asymptotic geometry never changes). This is a satisfying picture because it respects the target space quasi-locality property that world-sheet RG flow inherits from string theory.

At late RG ``times", we therefore have a non-linear sigma model whose target space geometry is a cone with an expanding smooth cap. In this regime, the correct tool for quantitatively following the flow is not the GLSM, but rather the one-loop sigma-model beta functions. The beta function equation for the metric is the Ricci flow equation:
\eqn\rgeqn{
{dG_{\mu\nu}\over d\lambda} =
-\alpha'R_{\mu\nu} + \nabla_\mu\xi_\nu +\nabla_\nu\xi_\mu,
}
where $\lambda=-\ln\Lambda$ is the RG ``time". The vector $\xi$ is arbitrary, representing the freedom to make continuous changes of target space coordinates along the flow. We need only concern ourselves with the flow of the metric, since the $B$-field must be trivial because we are in two dimensions. The dilaton may be doing something non-trivial, but inspection of the beta functions shows that the dilaton has no effect on the flow of the metric.

In principle solving this equation requires knowing the initial conditions, that is the geometry of the target space at the value of $\lambda$ where string effects first become negligible. However, since Ricci flow is diffusive in character, we would not actually expect the late time behavior to depend very much on the details of those initial conditions. Consider for example the heat equation, another diffusive first-order PDE. It has the property that for any initial conditions (with suitable asymptotics), after a sufficiently long time the distribution will become approximately Gaussian. Why a Gaussian rather than some other function? Because a Gaussian is the only function that retains its shape under diffusion: over time it simply becomes a broader and flatter Gaussian. We can similarly expect that, regardless of the details of the initial conditions, the target space will after some time evolve into a geometry that changes under Ricci flow \rgeqn\ only by an overall scaling (it is important that the conical asymptotic geometry is self-similar, otherwise this would not be possible).

There is in fact a unique smooth, rotationally symmetric two-dimensional geometry that is asymptotically conical with a given deficit angle, and that changes under Ricci flow only by an overall rescaling. It is \GuHeMiSc
\eqn\solone{
ds^2 = \lambda\left(f^2dr^2+{r^2\over n^2}d\theta^2\right), \qquad
\xi_r = {1\over2}rf,
}
or by making the $\lambda$-dependent change of coordinates $r = \rho/\sqrt{\lambda}$,
\eqn\soltwo{
ds^2 = f^2d\rho^2+{\rho^2\over n^2}d\theta^2, \qquad
\xi_\rho = {1\over2\lambda}\rho f(1-f).
}
The function $f$ interpolates smoothly between $f=1/n$ at $r=0$ and $f=1$ at $r=\infty$:
\eqn\fdef{
f = \left[1+W\left((n-1)\exp\left(n-1-{r^2\over2\alpha'}\right)\right)\right]^{-1},
}
where $W$ is the product log function (the inverse function of $xe^x$).

The RG flow parameter $\lambda$ in this solution ranges only from $0$ to $\infty$. The limits $\lambda\to0$ and $\lambda\to\infty$ are best taken at fixed $\rho$ (rather than fixed $r$), giving respectively the cone and the plane. For finite $\lambda$, as shown in fig.\ 2, the geometry is conical at infinity but smooth at the origin, with typical radius of curvature $\sqrt{\lambda\alpha'}$. Thus the curvature which was initially concentrated at the origin eventually diffuses over an infinite area. Note that we can only trust this solution for $\lambda\gg1$, when the curvature is 
small enough for eq.\ \rgeqn\ to be valid, so there is no significance to the fact that we cannot continue this solution to negative $\lambda$.

\fig{Cross section of an embedding in ${\bf R}^3$ of the geometry \solone, \soltwo\ in the case $n=3$.}{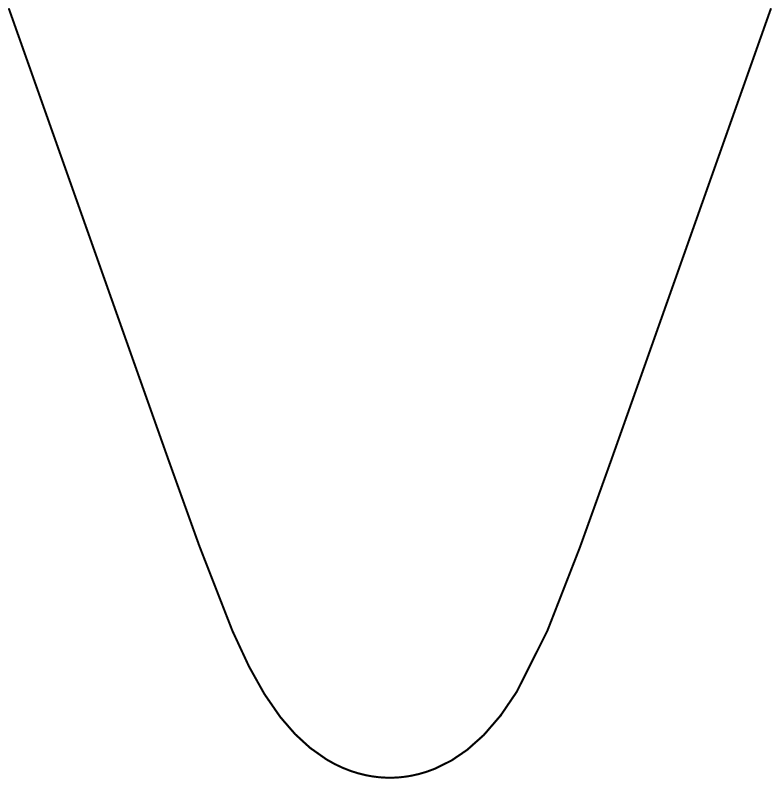}{2.2truein}

In the previous subsection, we saw that the flow seeded by the tachyon vertex operator $V_k$ leads to the lower-order orbifold ${\bf C}/{\bf Z}_k$. The solution described in this subsection can be generalized to describe such a flow simply by replacing $n$ by $n/k$ in the definition of $f$ \fdef.

\subsec{Time evolution: substringy regime}

We now turn to the classical spacetime dynamics of tachyon condensation in ${\bf C}/{\bf Z}_n$. In principle, the proper framework for studying these dynamics is closed string field theory. This subject is still in its infancy, however, and a direct attack remains a problem for the future. Instead, we shall find two opposite limits for which present tools can help us infer important qualitative features of the dynamics. In this subsection we will discuss the very early dynamics, before non-linear effects kick in, and in the next the very late dynamics, after stringy effects are washed away.

The classical dynamics of closed string fields does not depend on the string coupling, which appears in the action only in an overall prefactor. The supergravity actions are familiar examples of this, but it is true also for closed string tachyons such as the ones we are studying:
\eqn\gencsact{
S_T = {1\over\kappa^2}\int d^8\!x\left[
-\sum_{k=1}^{(n-1)/2}\left(\partial_\mu T_k\partial^\mu T_{-k} + m_k^2T_kT_{-k}\right) + L_{\rm int}
\right].
}
Here $L_{\rm int}$ includes the tachyons' interactions with each other as well as with the other closed string fields. The coefficients of the terms in $L_{\rm int}$ (as well as $m_k^2$) are of order 1 in string units. Therefore the linearized equation of motion for $T_k$, whose solution is
\eqn\eomsol{
T_k(x^0) = T_k(0)\cosh(|m_k|x^0) + {1\over|m_k|}\dot T_k(0)\sinh(|m_k|x^0),
}
is valid as long as $T_k^2\ll1$ and $\alpha'\dot T_k^2\ll1$. This will be true for a time that is logarithmic in $T_k(0)$ and $\dot T_k(0)$, so if we tune the initial values and velocities to be very small, then \eomsol\ will be valid over many string times.

Simply writing down the solution \eomsol, however, is not very satisfying. We really want to know what is happening to the spacetime while the tachyon field is growing according to \eomsol. How is its geometry changing? This is not a sensible question from the point of view of the string: the world-sheet theory is the ${\bf C}/{\bf Z}_n$ orbifold theory perturbed by the marginal operators $T_k(x^0)V_k$, which is simply not a sigma model. Luckily, there is another object in the theory we can use to probe the geometry: the D-brane. D-branes are elementary yet very massive (at weak coupling), and therefore good at probing very short spacetime distances \DoKaPoSh. The theory on a D-brane localized in a space such as ${\bf C}/{\bf Z}_n$ reduces at low energies to a moduli space, and the geometry of that moduli space is how the brane ``sees" the space. In the case of ${\bf C}/{\bf Z}_n$, the moduli space is simply ${\bf C}/{\bf Z}_n$ itself, as we will show below. However, when the tachyon field is turned on, the theory gets perturbed and as a result the moduli space is deformed. We will see that it remains asymptotically conical, but with the singularity replaced by a smooth cap whose area is proportional to the value of the tachyon field. We will merely sketch the calculations here; details can be found in the papers \refs{\AdPoSi,\chicago}.

The theory on a D-brane in ${\bf C}/{\bf Z}_n$ is the theory on $n$ D-branes in ${\bf C}$, projected down to the ${\bf Z}_n$-invariant states (there are no twisted sectors, because we are dealing with open strings). In the massless sector, the result is a $U(1)^n$ GLSM with $n$ complex transverse scalars and $n$ fermions. (If the brane has other transverse directions besides the ${\bf C}/{\bf Z}_n$ ones, then there will also be other transverse scalars, but we are not interested in those.) Each scalar is a bi-fundamental under a different pair of $U(1)$s. The pattern of charges is best summarized by the quiver diagram shown in fig.\ 3. Each node $j$ represents a different $U(1)$, whose gauge field we will denote $A_{jj}$. (The indices $j$ are defined mod $n$.) Each arrow, connecting a node $j$ to a neighboring one $j+1$, represents a scalar $Z_{j,j+1}$, which has charge $+1$ under $U(1)_j$ and $-1$ under $U(1)_{j+1}$. The fermions are also bi-fundamentals, but with a different quiver diagram.

\fig{Quiver diagram for the scalars of the ${\bf C}/{\bf Z}_n$ probe theory.}{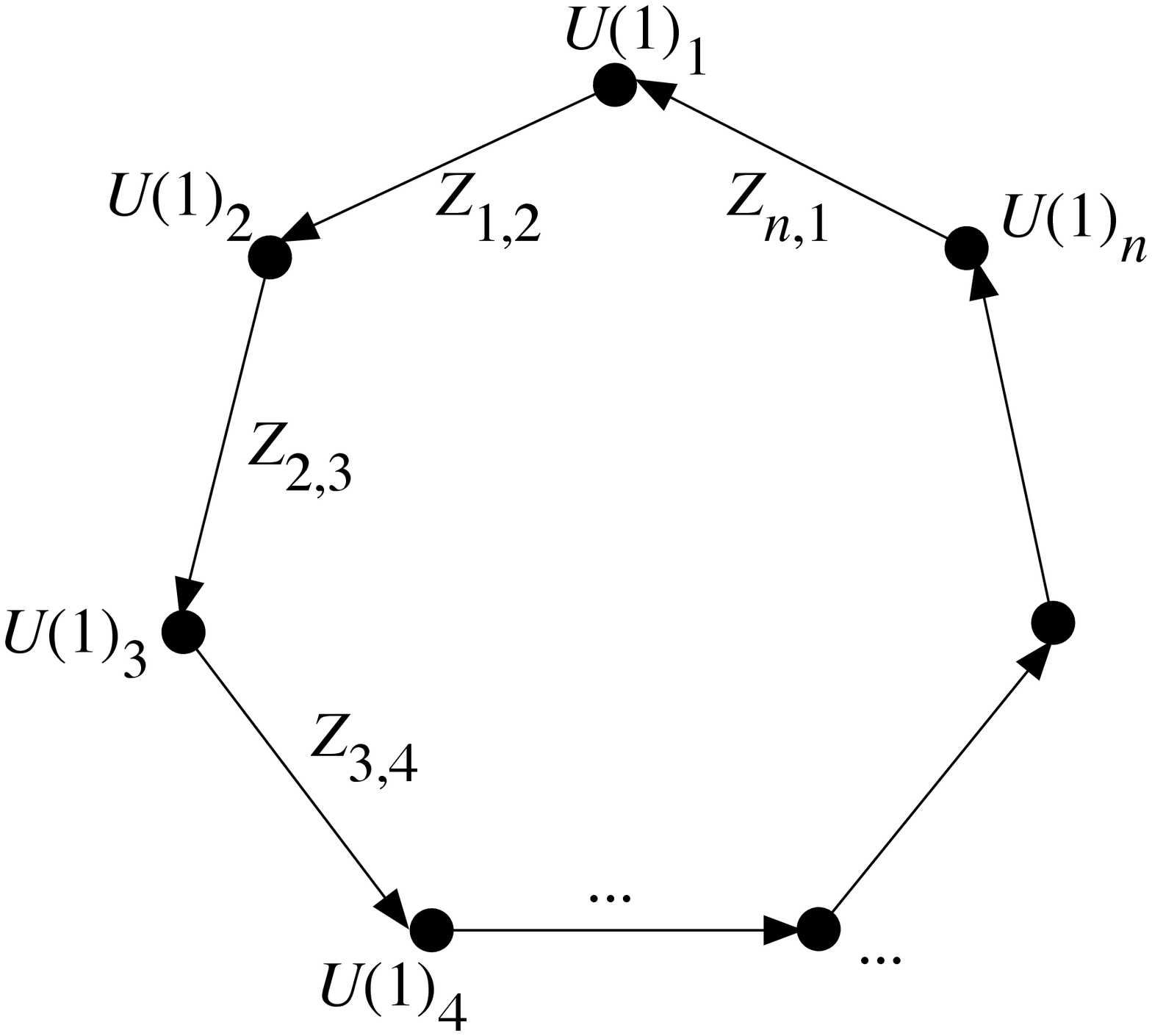}{3.0truein}

Unlike the theories we studied in subsection 3.2, this GLSM is non-supersymmetric: supersymmetry has been broken by the orbifold projection, just as it was in the closed string theory. However, its classical action is still determined by supersymmetry, since it is simply an orbifolded version of a supersymmetric theory (and there are no twisted sectors). In particular, in addition to their gauge interactions, the scalars have a D-term potential:
\eqn\dtermpot{
V = {1\over2}\sum_{j=1}^n\left(|Z_{j,j+1}|^2-|Z_{j-1,j}|^2\right)^2.
}
To explain the derivation of the moduli space from this potential, we can do no better than to quote the original reference \AdPoSi\ (subsection 2.2):
\smallskip{\narrower\noindent\baselineskip = 15 pt
The vanishing of the potential [\dtermpot] implies that the magnitude $Z_{j,j+1}$ is independent of $j$. Of the $n$ $U(1)$ symmetries, the diagonal decouples. The remaining $n-1$ gauge symmetries can be used to set the phases of the $Z_{j,j+1}$ equal as well, so that the common value $Z_{j,j+1}=Z$ parametrizes the branch. The branch is thus two-dimensional, as it should be for the interpretation of a probe. The gauge choice leaves unfixed a ${\bf Z}_n$ gauge symmetry, whose generator is $\exp(-2\pi i\sum_jjQ_j/n)$. This identifies $Z\to e^{2\pi i/n}Z$, so the probe moduli space is indeed the ${\bf C}/{\bf Z}_n$ spacetime.

}\noindent
The authors then proceed to check that the metric on this moduli space, obtained by integrating out the massive gauge bosons, is indeed that of ${\bf C}/{\bf Z}_n$.

Singularities in moduli spaces are typically associated with points of enhanced gauge symmetry, and this case is no exception: at the origin all of the $Z_{j,j+1}$ vanish simultaneously and the $U(1)^{n-1}$ gauge symmetry is restored. The $Z_{j,j+1}$ are related to each other by the ${\bf Z}'_n$ quantum symmetry of the orbifold, which manifests itself in the probe theory as the global symmetry generated by the permutation $j\to j+1$ (mod $n$), i.e.\ the discrete rotational symmetry of the quiver diagram (fig.\ 3). The ${\bf Z}'_n$ symmetry is thus responsible for the fact that the $Z_{j,j+1}$ all vanish at the same point on the moduli space, and therefore for the conical singularity appearing at that point. By this logic, we would expect that turning on tachyons and breaking the symmetry should make the singularity go away.

To test this conclusion, we should see how the operator $\sum_kT_kV_k$ added to the world-sheet Lagrangian changes the open string dynamics on the D-brane, in particular the low energy gauge theory. This was done by the authors of \chicago, who calculated the disk diagram with two scalar vertex operators on the boundary and a tachyon vertex operator in the interior. They found that the scalar potential \dtermpot\ is changed to quadratic order by the addition of a term
\eqn\deltaV{
\Delta V =
- \sum_j\zeta_j\left(|Z_{j,j+1}|^2 - |Z_{j-1,j}|^2\right),
}
where $\zeta_j$ are the discrete Fourier transform of the tachyon vevs $T_k$,\foot{The reader may object that we are treating the $T_k$ as constants, whereas in fact they are world-sheet operators because they depend on $x^0$. In this sense the calculation that follows should be taken as providing heuristic support for the smoothing out of the cone, since in practice the $\zeta_j$ parameters will be varying in time too fast for the moduli space approximation to be valid. One way around this problem is to give the tachyons not a time- but a space-dependence, for example $T_k(x^1) = T_k(0)\cos(|m_k|x^1)$. A D-brane localized in the $x^1$ direction would be able to resolve this $x^1$ dependence, and its moduli space would be a fibration over the $x^1$ direction of the one we find for the orbifold directions.}
\eqn\zetadef{
\zeta_j = \frac C{2i}\sum_{k=1}^{n-1}e^{-2\pi ijk/n}(-1)^kT_k,
}
with $C$ is a real constant. (Note that the relation $T_k^* = T_{n-k}$ guarantees that the $\zeta_j$ are real.) Adding \deltaV\ to \dtermpot\ deforms the ``D-term constraint" to be
\eqn\newdterm{
|Z_{j,j+1}|^2 - |Z_{j,j-1}|^2 = \zeta_j.
}
In particular, for generic tachyon vevs all the $Z_{j,j+1}$ must be different. If only one of them can vanish at a time, then there will be no point of enhanced symmetry on the moduli space and therefore no singularity. Indeed, the authors of \AdPoSi\ showed this to be the case by integrating out the gauge bosons and finding the metric
\eqn\modmet{
ds^2 = n(r)dr^2 + {r^2\over n(r)}d\theta^2.
}
Here $n(r)$ is a function that depends on the $\zeta_j$ parameters; when they vanish it equals $n$ identically, but for generic values of the $\zeta_j$ it interpolates smoothly between $n(0)=1$ and $n(\infty)=n$. The curvature radius of the smooth cap goes like the square root of the $\zeta_j$ parameters, and hence of the value of the tachyon fields.

\subsec{Time evolution: gravity regime}

What happens during the stringy phase of the decay, when the tachyon fields $T_k$ and their time derivatives are string scale? Lacking the tools for a quantitative analysis in this regime, let us see how far we can get by general reasoning. First we recall that, from the spacetime point of view, the orbifold fixed point is a localized positive-energy object, a soliton of the closed string field theory. Furthermore, since gravity does not act as a binding force on co-dimension 2 objects, it is not surprising that this configuration should be unstable (besides the fact that it's non-BPS). When the soliton decays, nothing stops its energy from diffusing outward (barring a cataclysmic event such as a the formation of a tear in spacetime). The energy density is at first so high that we may expect it to be transformed into very massive strings, as occurs in the decay of D-branes discussed in subsection 2.4. These strings will eventually decay into massless ones that fly away from the origin at the speed of light. When this happens, and once the energy densities and curvatures are below the string scale, then the low energy effective theory---supergravity---becomes an appropriate description. We re-emerge into the light, a regime where a quantitative analysis is once again possible.

In this subsection we will study the supergravity equations of motion for this system, and find that they are surprisingly tractable, particularly considering the fact that we are dealing with a time-dependent, non-supersymmetric process. The discussion summarizes the paper \He, which should be consulted for details; see also \GrHa. We first show that the dynamics reduces to a massless scalar, the dilaton, minimally coupled to $2+1$ gravity. Recalling that gravitons cannot go on-shell and carry energy in $2+1$ dimensions, we see that the dilaton plays an indispensible role, namely to carry the energy released in the decay off to infinity. Usually in general relativity such coupled matter-gravity equations of motion are very difficult to solve, due to the problem of gravitational back-reaction. However, we will see in this case by using the rotational symmetry and making a judicious choice of coordinate system that the equation of motion for the dilaton decouples from the back-reaction of the geometry: it is the same as that for a dilaton in Minkowski space. This allows the equations to be solved in two steps: first, the dilaton equation of motion is straightforwardly solved, then the back-reaction of the geometry is found by solving a first-order equation for a metric coefficient that is sourced by the dilaton's energy. Among other things, this method guarantees that under non-singular initial conditions the energy will indeed radiate to infinity and no singularity can form.

Let us begin by showing that the dilaton and metric are the only fields we need to consider. Since the tachyons are all in the NS-NS sector, we can consistently truncate to that sector (remember, we are considering a classical process). We can also consistently truncate the dynamics to $2+1$ dimensions, namely the two of the orbifold plus time \AdPoSi. (The CFT describing the extra seven dimensions is coupled to the $2+1$ dimensional CFT neither in the initial configuration nor by the tachyon vertex operator, and therefore decouples entirely.) Finally, the equation of motion for the $B$-field requires its field strength to be a constant multiple of the volume form on the unreduced $2+1$ dimensions. This constant vanishes in the initial orbifold configuration, and must therefore vanish everywhere.

The Einstein equation can be written
\eqn\einstein{
R_{\mu\nu} = 4\partial_\mu\Phi\partial_\nu\Phi
}
(we are working in Einstein frame). By rotational invariance the $\theta$--$\theta$ component of this equation is $R_{\theta\theta}=0$, and one can show that the general solution is a metric of the form
\eqn\fixed{
ds^2 = e^{2\sigma(t,r)}(-dt^2+dr^2) + r^2d\theta^2.
}
The static orbifold is described by a constant value of $\sigma$:
\eqn\static{
\sigma = \ln n \qquad ({\bf C}/{\bf Z}_n).
}
The great advantage of the metric \fixed\ is that the unknown function $\sigma$ drops out of the dilaton's equation of motion (assuming again that it doesn't depend on $\theta$); just as in flat space, we have
\eqn\scalars{
\left(\partial_t^2 - \frac1r\partial_rr\partial_r\right)\Phi = 0.
}
This linear equation can be straightforwardly solved using standard methods. One implication of it is that, regardless of what happens during the decay, the dilaton will at late times return to its original value. Given a solution to \scalars, the back-reaction of the dilaton's energy on the geometry can be determined by the remaining components of the Einstein equation \einstein:
\eqn\sigmaeomtr{
\partial_t\sigma = 4r\partial_t\Phi\partial_r\Phi, \qquad
\partial_r\sigma = 2r\left((\partial_t\Phi)^2+(\partial_r\Phi)^2\right).
}

As an example, fig.\ 4 shows a numerical solution (obtained using {\it Mathematica}) for the decay of ${\bf C}/{\bf Z}_3$ to the plane. The decay occurs at $t=0$, and in order to simulate the energy being dumped from the defect into the dilaton, a source term was added to the right hand side of \scalars:
\eqn\exsource{
\left(\partial_t^2 - \frac1r\partial_rr\partial_r\right)\Phi  = \beta e^{-t^2-r^2},
}
where the normalization $\beta=1.2$ was set by requiring $\sigma$ to decrease by $\ln 3$ from $t=-\infty$ to $t=\infty$. We see that the outgoing dilaton wave carries with it the cone's curvature, leaving behind an expanding flat disk. Note that the geometry at late times is qualitatively different from that at late ``times" in the world-sheet RG flow (fig.\ 2), which is much rounder, reflecting the difference between the first- and second-order equations governing the two different types of evolution.
\fig{Numerical simulation of decay of ${\bf C}/{\bf Z}_3$ to the plane, using the source function \exsource. Each of the four figures shows a constant $t$ slice, at $t=-3$ (top left), $t=3$ (top right), $t=9$ (bottom left), and $t=15$ (bottom right). For each $t$ the field configuration $\Phi(t,r)$ is plotted above a cross-section of an embedding in ${\bf R}^3$ of the geometry.}{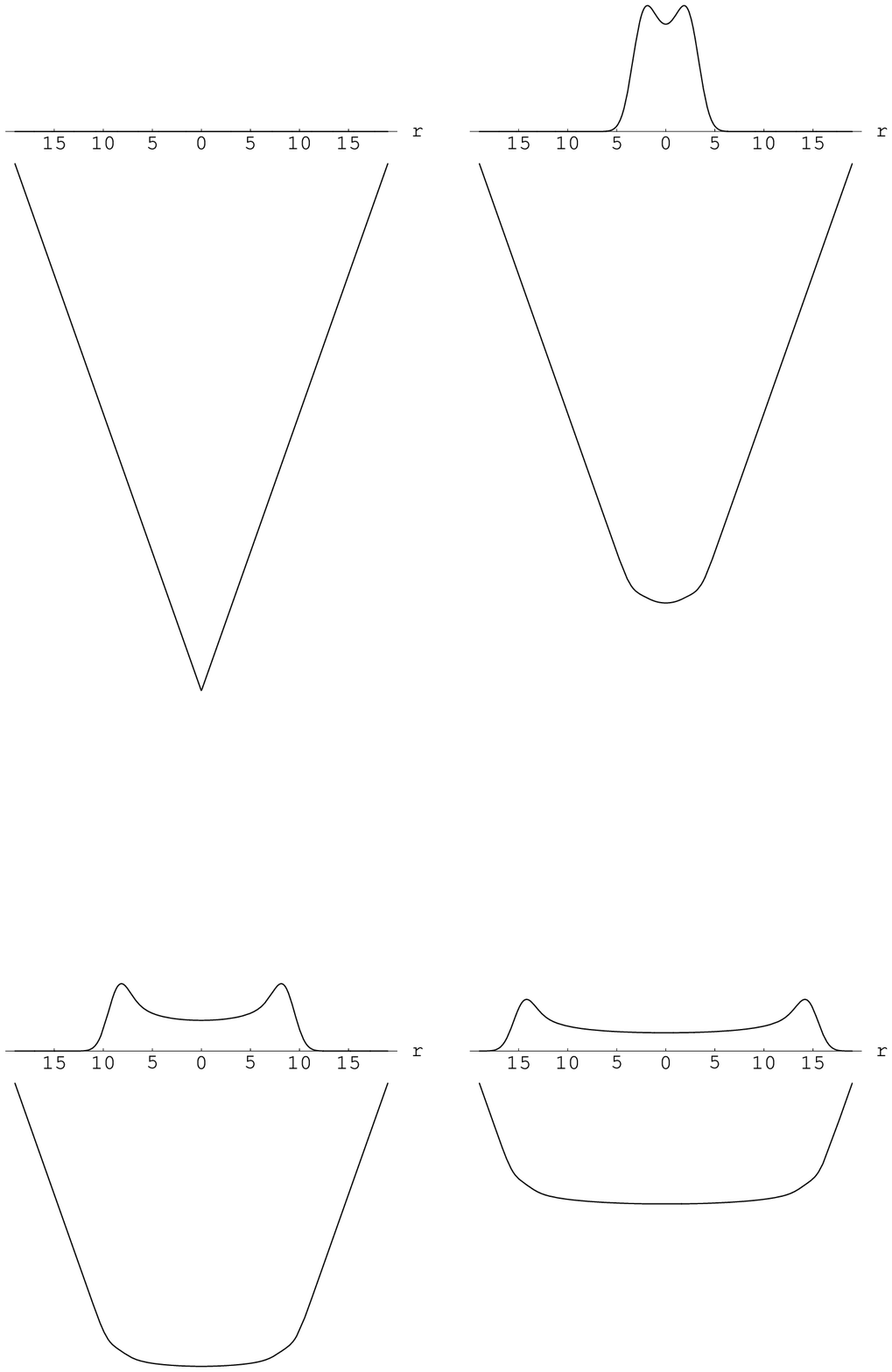}{4.2truein}

\newsec{Decay of Twisted Circles}

\subsec{Introduction}

From the example of the ${\bf C}/{\bf Z}_n$, it might appear that in order to localize closed string tachyons it is necessary to have some singularity in the spacetime, a defect on which the tachyons live. In fact this is not the case, as shown by the twisted circle compactifications that are the subject of this section. The twisted circle is an SCFT with a three dimensional target space, constructed by twisting the free $\hat c=3$ sigma model by a geometrical symmetry of its ${\bf R}^3 = {\bf C}\times{\bf R}$ target space. Whereas twisting by a discrete translation $e^{2\pi iRp_y}$, where $y$ is the coordinate in the ${\bf R}$ direction, would produce the usual circle compactification (times the trivial plane ${\bf C}$), the twisted circle is obtained by combining this translation with a rotation of the plane:
\eqn\twistact{
e^{2\pi i(\zeta J+ R p_y)},
}
where $J$ is the angular momentum in the ${\bf C}$ direction and $\zeta$ is a real parameter which is defined mod 2, $\zeta\sim\zeta+2$. This combined symmetry transformation acts freely on ${\bf C}\times{\bf R}$; modding out by it leaves a non-singular geometry that should be thought of as a fibration of a plane over a circle: on going around the circle, the plane gets rotated through an angle $2\pi\zeta$. Spacetime supersymmetry is broken as long as $\zeta\neq0$ (mod 2).

With two parameters, the circle radius $R$ and the twist parameter $\zeta$, this model is fairly versatile, reducing in special cases other interesting models. For example, at $\zeta=1$ the symmetry \twistact\ becomes $(-1)^Fe^{2\pi iRp_y}$. In this case the plane ${\bf C}$ factors out trivially, but the circle has anti-periodic boundary conditions for fermionic fields; this type of circle goes by the various names, including Scherk-Schwarz compactification, Rohm compactification, thermal circle, and interpolating orbifold. Below we will see that, in a different limit, the twisted circle reduces to the ${\bf C}/{\bf Z}_n$ orbifolds studied in the previous section.

At generic $R$ and $\zeta$, the twisted circle is also dual to some other interesting spacetimes. For example, Kaluza-Klein reduction along the $U(1)$ isometry generated by $\zeta J+Rp_y$ yields a Melvin configuration of the Kaluza-Klein gauge field. On the other hand, T-dualizing along this isometry yields a co-dimension two flux-brane of the NS-NS 3-form flux.

The spectrum of type II strings on the twisted circle is straightforward to derive \refs{\tse,\TseytlinEI,\TseytlinZV} (see \refs{\TaUeo,\DaGuHeMi} for a brief review). The lightest twisted field has a mass given by
\eqn\twistmass{
m^2 = {R^2\over{\alpha'}^2} -  {2|\zeta|\over\alpha'} \qquad (|\zeta|\le1)
}
(the restriction on $\zeta$ is for notational convenience; since $\zeta\sim\zeta+2$, it implies no loss of generality). Thus the system is perturbatively unstable if and only if
\eqn\twistunstable{
|\zeta| > {R^2\over2\alpha'}.
}

Even when the theory is perturbatively stable, however, as long as supersymmetry is broken it will have a non-perturbative instability. For example, the Scherk-Schwarz compactification ($\zeta=1$) was shown by Witten \WiBON\ to decay via a bounce to nothing at all. The non-perturbative decay of the Melvin dual to the twisted circle by nucleation of branes has also been studied \refs{\EmGu,\Gu,\AharonyCX}, but there is disagreement in the literature about the results.

When the twist parameter is rational, we can express it as $\zeta=2m/n$, with $n>0$ and $m$ and $n$ relatively prime. In this case the twisted circle can be constructed by a ${\bf Z}_n$ twist of ${\bf C}\times S^1_{nR}$, where $S^1_{nR}$ denotes a normal circle of radius $nR$. This is convenient, since it means we have only a finite number ($n-1$) of twisted sectors. If $n$ is even (and therefore $m$ is odd) then the orbifold group includes $(-1)^Fe^{\pi inRp_y}$. The strings twisted by this element see the space as a Scherk-Schwarz circle of radius $nR/2$ times a plane, and are therefore delocalized, just as in the case of ${\bf C}/{\bf Z}_n$ at even $n$. We will therefore take $n$ odd; the result is that strings in all twisted sectors are localized, since they must stretch in order to move away from the origin of the plane. For simplicity we will restrict ourselves further in this section to the case $m=(n+1)/2$, i.e.\ to the orbifold generated by
\eqn\twistidn{
(-1)^Fe^{2\pi i(J/n+Rp_y)}.
}

This special case of the twisted circle is still versatile enough to have some interesting limits. In the limit $n\to\infty$ with $R$ fixed, it goes over to the Scherk-Schwarz circle times the plane. On the other hand, in the limit $R\to0$ with $n$ fixed, it goes over to our familiar ${\bf C}/{\bf Z}_n$. The way to see this is to note that the original circle (of radius $nR$) is becoming very small. A T-duality yields a circle of radius $\alpha'/nR\to\infty$ times the plane. Only strings with zero winding, i.e.\ $p_y=0$, survive, so the orbifold action \twistidn\ is reduced to $(-1)^Fe^{2\pi iJ/n}$, and we have ${\bf C}/{\bf Z}_n\times{\bf R}$. Finally, if we simultaneously take $n\to\infty$ and $R\to0$ holding $nR$ fixed, we end up, after a T-duality with respect to $J/n+Rp_y$, with an $H$ flux-brane in type 0 theory.

It is reasonable to conjecture by analogy with the ${\bf C}/{\bf Z}_n$ system that the condensation of the tachyons in this orbifold leads to the orbifold being undone, leaving the supersymmetric ${\bf C}\times S^1_{nR}$ background. The main evidence for this conjecture comes not from time evolution or world-sheet RG flow, but from a process that can be considered somewhere in between the two, namely Liouville flow \KutasovPF. Following \DaGuHeMi, in the next subsection we will construct a CFT with a four-dimensional target space, one of whose dimensions is a Liouville direction. (See also \refs{\SuyamaGD,\piljin} for different approaches to the same problem.) The other three directions are fibered over this Liouville direction in such a way that in one limit---the ``UV" end---they form the twisted circle, and at the other limit---the ``IR" end---they are simply ${\bf C}\times S^1_{nR}$. In subsection 4.3 we will take a closer look at how the geometry changes along the flow---which cycles disappear, and how. Finally, in subsection 4.4 we will use this information to deduce what happens to stable branes, both regular and fractional, when the background they are living on decays.

\subsec{GLSM construction of the Liouville flow}

The Liouville flow CFT is straightforwardly constructed using the GLSM techniqe. We will sketch it here; for details, see \DaGuHeMi. The GLSM we need is very similar to the one we used in subsection 3.2. It has a $U(1)$ gauge group and two chiral superfields $\Ph_1$ and $\Ph_{-n}$ with charges 1 and $-n$ respectively. However, in addition it has an axionic field $P=P_1+iP_2$. The target space for this field is a cylinder, $P\sim P+2\pi i$, and under a gauge transformation with gauge parameter $e^{i\alpha}$ it is translated rather than rotated:
\eqn\ft{
P \to P + i \alpha.
}
The action for this system is
\eqn\liouvaction{
S = {1\over 2\pi}\int d^2\!\sigma d^4\theta
\left[
\bar\Phi_1e^V\Phi_1 + \bar\Phi_{-n}e^{-nV}\Phi_{-n} + {k\over4}(P+\bar P+V)^2 - {1\over2e^2}|\Sigma|^2
\right],
}
where $k$ is a free parameter that will determine the size of the $S^1$. As in the GLSM of subsection 5.1, the low energy dynamics is described by a sigma model onto the manifold of supersymmetric zero-energy configurations, modulo gauge equivalences. The D-term condition is
\eqn\zet{
-\f{D}{e^2} = |\ph_1|^2-n|\ph_{-n}|^2+k P_1=0.
}
Note that $P_1$ plays the role of a dynamical FI term. (As in subsection 3.2, in order to impose the type II GSO projection on the low energy theory we should mod this GLSM out by an appropriate ${\bf Z}_2$ R-symmetry.)

At positive values of $P_1$, $\ph_{-n}$ must get a vev, so the gauge group is Higgsed down to ${\bf Z}_n$. Since $\ph_1$ parametrizes ${\bf C}$ and $P_2$ parametrizes an $S^1$ of radius $\sqrt{k\alpha'}$, the target space is the twisted circle $({\bf C}\times S^1_{nR})/{\bf Z}_n$, with $R = \sqrt{\alpha'k}/n$. That the metric is indeed that of $({\bf C}\times S^1_{nR})/{\bf Z}_n$ can be checked as usual by integrating out the gauge bosons \MiTa. This metric will only be valid at large $P_1$, since then the gauge bosons are very massive and quantum effects are suppressed. On the other hand, when $P_1$ is negative it is $\ph_1$ that gets a vev, and the gauge group is completely broken. The target space here is simply ${\bf C}\times S^1_{nR}$, with this ${\bf C}$ parametrized by $\ph_{-n}$. Again, the metric can be checked.

In \DaGuHeMi\ it was argued that $P_1$ may be regarded as a Liouville or world-sheet scale direction, with $P_1 \r \infty$ corresponding to the UV and $P_1\r-\infty$ to the IR. Consequently, the change in the transverse geometry as $P_1$ goes from infinity to minus infinity mimics their flow as the energy scale goes from the UV to the IR. Another way of thinking about it is that the coordinate $-P_1$ can be regarded as the real time $t$ after a Wick rotation. We conclude that the result of tachyon condensation in the twisted circle is that the ${\bf Z}_n$ orbifold is lifted, and the theory decays into a normal, supersymmetric circle compactification.

What does this result say about the various limits described in the last subsection? If we take $n\to\infty$ with $R$ fixed, then we would say that a Scherk-Schwarz circle decays perturbatively to non-compact space. Taking $R\to0$ holding $n$ fixed, we find that ${\bf C}/{\bf Z}_n$ also decays to flat space, in agreement with the findings of the previous section. Finally, taking $n\to\infty$ and $R\to0$ holding $nR$ fixed, we find that the type 0 flux-brane decays to a supersymmetric circle in type II theory (more precisely, 0A would decay to IIB and 0B to IIA).

\subsec{Classical geometry of the Liouville flow}

Even though we have understood the beginning and final
geometry of the Liouville flow in the previous subsection,
it is interesting to examine its middle
process in detail. This will give us an intuitive idea how
the spacetime geometry will change under closed string tachyon 
condensation. In this subsection we will describe the effective 
geometry of the Liouville flow in some detail. 

It is not difficult to compute the ``classical" effective metric
(sigma model metric) implied by the GLSM construction 
to the leading order of stringy corrections.
In the first step we plug the D-term equation \zet\ into the GLSM action,
fix the $U(1)$ gauge (by setting the phase of $\ph_{-n}$ to 1)
and then integrate out the (massive) gauge field which
appears quadratically in the Lagrangian. These manipulations are easy to
perform, and lead in the sigma model limit
$e \to \infty$ to an ${\cal N}=(2,2)$ sigma model
on a smooth four dimensional target space \MiTa
\eqn\metrico{ds^2=  d \rho^2+ d \rho'^2
+\f{2}{k}(n\rho' d \rho' -\rho d\rho)^2 +
 {{k \over 2} \rho^{2}d\tilde{\theta}_1^2
 +{k \over 2}\rho'^2d\tilde{\theta}_2^2
 +\rho^2\rho'^2
 (nd\tilde{\theta}_1+d \tilde{\theta}_2)^2
 \over
\rho^2 + n^2 \rho^{'2}+{k \over 2}  }.}
The coordinates that appear in \metrico\ are defined by
$\rho=|\ph_1|,  \rho'=|\ph_{-n}|, \tilde{\theta}_1= \theta-P_2$ and
$\tilde{\theta}_2=nP_2$, where $\theta$ is the angle of $\ph_1$.
Then $\tilde{\theta}_{1,2}$
are periodic with periodicity $2\pi$.

The (Liouville) conformal field theory that captures the process of 
tachyon condensation of the twisted circle theory may now be obtained 
by flowing to the IR in the world-sheet sigma model based on the metric
\metrico. Unfortunately, an exact description of the
resulting ${\cal N}=(2,2)$ SCFT has not yet been found; determining 
this SCFT is an interesting open problem  (note \HoKa\ were able to 
solve a similar problem in a slightly simpler context; the 
CFT that described the IR of their model was the $SL(2,R)/U(1)$ WZW theory).
It seems plausible that, as in \HoKa\ (and several earlier analysis of Calabi
Yau constructions---see for instance \WittenYC), the asymptotic geometry
and topology contained in the exact SCFT match those of \metrico. Assuming 
this is the case, we now proceed study the qualitative features of the 
geometry  \metrico\ in detail (we refer to \metrico\ as the ``classical"
geometry of tachyon condensation).

\fig{Geometry change under the closed
string tachyon condensation. The bubble of flat space appears at the origin
when $P_1=0$. Later it will expand and finally it covers the whole space.}
{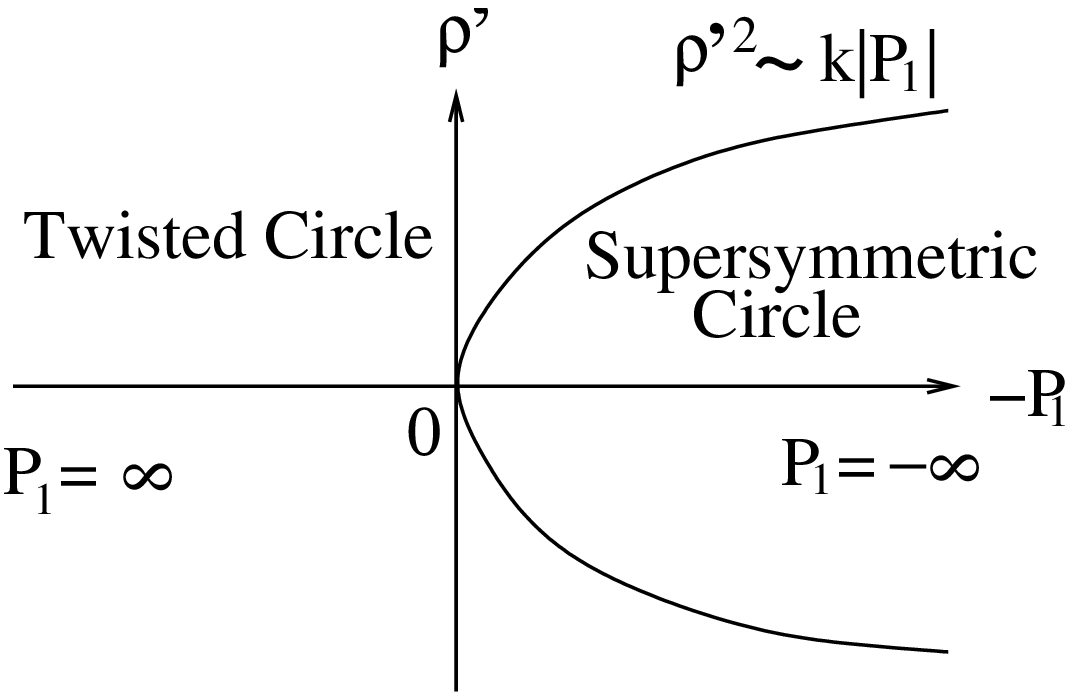}{3.0truein}

As we have seen before, the geometry
\metrico\ should describe a space that interpolates
between the twisted circle geometry and flat space. It is easy to see
this explicitly from the metric \metrico. 
Rewriting the metric in terms of the coordinates
$\rho$, $\theta$, $P_1$ and $P_2$,
we find that, in the UV limit ($P_1 \gg 0$) the interpolating space reduces to
\eqn\metricone{\eqalign{
ds^2&=(d\rho)^2+\f{k}{2}(dP_1)^2+\f{(\rho d\rho
+\f{k}{2}dP_1)^2}{n(\rho^2+kP_1)}\cr
&{}\qquad+\f{(n\rho^4+(nkP_1\!+\f{k}{2})\rho^2)(d\theta)^2
-k\rho^2(d\theta)(dP_2)+\f{k}{2}((n+1)\rho^2+nkP_1)(dP_2)^2}
{(n+1)\rho^2+nkP_1+\f{k}{2}}\cr
& \approx (d\rho)^2+\rho^2(d\theta)^2
+\f{k}{2}(dP_1)^2+\f{k}{2}(dP_2)^2.\cr}}
This indeed represents the twisted circle since the unbroken gauge
symmetry ${\bf Z}_n$ leads to the identification
$(\ph_1,P_2)\sim(e^{2\pi i/n}\ph_1,P_2+2\pi i/n)$.

On the other hand, rewriting
the metric in terms of the coordinates $\rho'$, $\theta$, $P_1$, and
$\tilde{\theta_1}$, we see that the interpolating space reduces to flat
space (i.e.\ supersymmetric circle) in the IR limit ($P_1 \ll 0$)
\eqn\metrictwo{\eqalign{
ds^2 &= (d\rho')^2+\f k2(dP_1)^2+\f{(n\rho' d\rho'+\f{k}{2}dP_1)^2}{n\rho'^2+kP_1}\cr
&\qquad{}+\f{
\f12k(n\rho'^2+kP_1)(d\tilde{\theta}_1)^2+n^2\rho'^2(n\rho'^2+kP_1)d\theta^2+\f12kn^2\rho'^2(d\tilde{\theta}_1-d\theta)^2}
{n(n+1)\rho'^2+kP_1+\f12k}\cr
&\approx d\rho'^2+n^2\rho'^2 (d\theta)^2+\f{k}{2}(d\tilde{\theta}_1)^2+\f{k}{2}(dP_1)^2,
}}
taking the ${\bf Z}_n$ identification into account. 

In greater detail, the geometry 
\metrico\ represents the decay of the twisted circle
background via the  nucleation and growth of a bubble of 
the (compactified) flat space. This bubble is nucleated at
``time" $P_1= 0$ (and at $\rho=\rho'=0$).  The radius $R_b$ of this
nucleated bubble (see fig.\ 5) grows in time
like $R_b\sim \sqrt{-kP_1}$ (this
diffusive growth\foot{In order to see this note that, in the far region $\rho'^2\gg k|P_1|$
\metrico\ reduces to the twisted circle, while in the opposite limit
$\rho'^2\ll k|P_1|$ we get the flat space. $\rho'^2 \approx k|P_1|$
(see fig.\ 5) represents a transition between these regions.} matches the behavior observed in section 3.2). These results
 precisely
show the expected behavior of the tachyon condensation
process.

In this way we have found the effective metric of the geometry
which interpolates between the twisted circle and flat space. An important
point compared with the previous RG flow description is that in
the present example the whole geometry corresponds to a CFT since
we have identified the Liouville coordinate $P_1$ with the ``time"
coordinate under the tachyon condensation. To check that our
derivation of the leading order effective metric is plausible, we
must show that our metric has no singularity. To examine this we
focus on the point at which the bubble of supersymmetric circle is
first nucleated, i.e.\ the neighborhood of $P_1=\rho=\rho'=0$. Then
\metrico\ reduces to \eqn\metricflat{ds^2=d\rho^2+\rho^2
d\tilde{\theta}_1^2 +d\rho'^2+\rho'^2 d\tilde{\theta}_2^2,} i.e.\
polar coordinates on ${\bf R}^4$. Thus there is no
singularity in our geometry.

We end this section by studying the essential topological properties of 
the geometry; this analysis will prove useful in the next section. 
In particular we will trace the behavior of various 1-cycles in the
twisted circle, as
the space transforms itself from the twisted circle to flat space. The
twisted circle $S^1_A$ defined by (see fig.\ 6)
\eqn\defcc{S^1_A:(\rho,\theta,P_2)=(0,s_1,s_1),\ \ 0\leq
s_1 \leq 2\pi/n,}
which is non-contractible for $P_1 >0$ (UV region), shrinks
and disappears at $P_1=0$. This is
simply the angle circle $\tilde{\theta_2}$ of one of the two planes in
\metricflat. This cycle becomes trivial for $P_1<0$.
\fig{One-cycles in the twisted circle.}{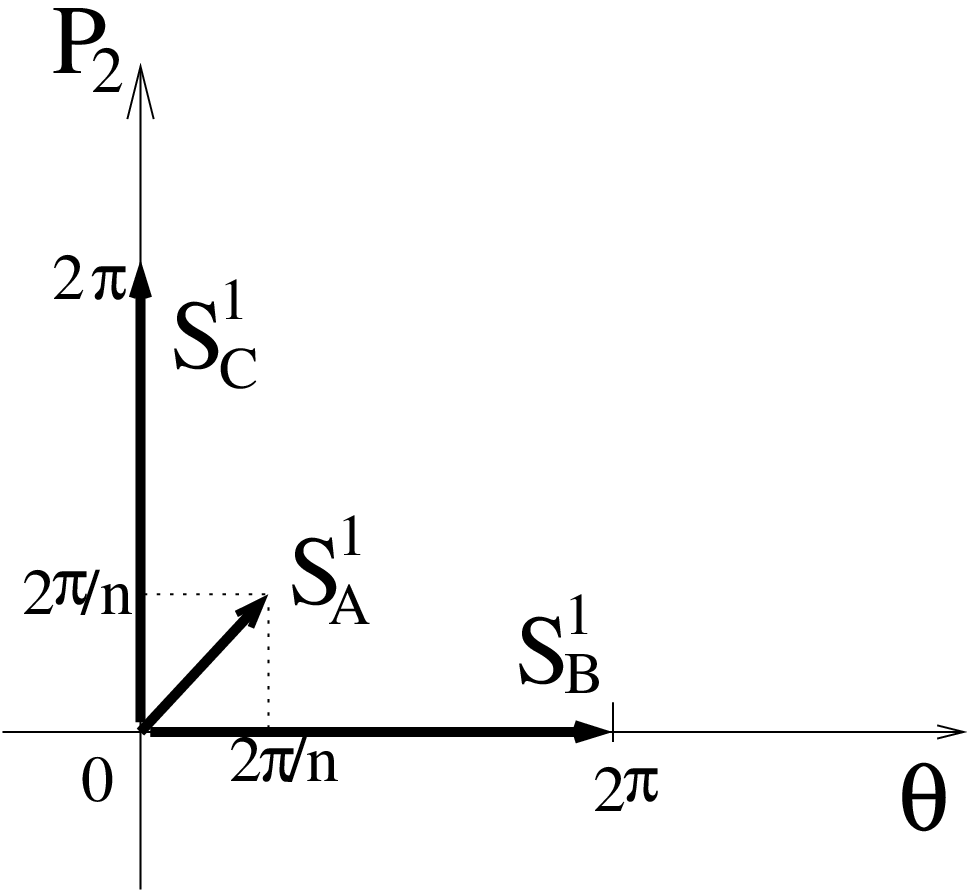}{2.5truein}

It is also interesting to track the evolution of the supersymmetric circle
$S^1_{B}$ defined by
\eqn\defthc{S^1_B:(\rho,\theta,P_2)=(0,s_2,0),\ \
0\leq s_2 \leq 2\pi,}
back in ``time", from the IR to the UV. This cycle (see fig.\ 6)
also vanishes at $P_1=0$; in fact it is simply the other angle
${\tilde \theta}_1$ in \metricflat. Thus $S^1_B$ 
is topologically non-trivial only for $P_1<0$ 
as opposed to $S^1_A$.

We can also consider the third cycle (see fig.\ 6)
\eqn\deftc{S^1_C:(\rho,\theta,P_2)=(a,b,u),\ \
 0\leq u \leq 2\pi,\ a,b={\rm constant}
 }
which winds $n$ times around the twisted circle in the UV ($P_1 \gg 0$) and
winds once around the supersymmetric circle in the IR ($P_1 \ll 0$). Note
that this cycle vanishes at the point $\rho=P_1=0$.

\subsec{Evolution of D-branes}

D-branes are very useful probes for capturing the geometry in string
theory, as we already discussed in section 3.4. In the twisted
circle as well as in ${\bf C/ Z}_n$ we have a rich D-brane
spectrum, including a bulk D-brane (regular D-brane) and a
fractional D-brane \refs{\DoMo,\DiDoGo} (see also \Bi\ for a review). 
The latter is special to
orbifold theories and is localized at their fixed point, while the
former has the same properties as usual D-branes in flat space.
Notice that they are all stable D-branes in unstable closed string
backgrounds. On the other hand, the endpoint of the tachyon
condensation, flat space, has just the simple D-brane spectrum. Thus
one may ask how the two different D-brane spectra are related to
each other via closed string tachyon condensation (or equally
RG-flow or Liouville flow).

Motivated by this, here we would like to study the evolution of
D-branes under tachyon condensation. We can concentrate on the
results in the twisted circle because the orbifold ${\bf C/ Z}_n$
can be obtained by the $R\to 0$ limit. In \MiTa\ two complementary
approaches are considered to study this problem.
(Refer to \piljin\ for an analysis of the 
world-volume theory on probe bulk D0-branes; see \refs{\MaMo,\MoP} 
for discussions on the fate of D-branes under the  
RG flow especially in the unstable orbifold ${\bf C}^2/{\bf Z}_n$ from
other viewpoints.) 
The first approach uses the important
fact that the Euclidean (more precisely Liouville) evolution of
twisted circle models is approximately captured by the smooth
interpolating geometry discussed in the previous subsection. 
We then note that D-branes on twisted circle backgrounds \refs{\DuMo,\TaUet}
may be thought of as usual type II branes wrapping warious 
geometric cycles of the 
twisted circle background \MiTa. It is then possible to follow the 
evolution of these cycles to late Euclidean ``time" in
the interpolating RG-flow geometry. It turns out that the cycles
wrapped by twisted circle D-branes always evolve under RG flow in
one of two distinct ways: the UV cycle either unambiguously 
evolves into an IR cycle, or it disappears at a finite ``time" 
(perhaps reappearing after that time). This leads to a variety 
of possibilities for D brane evolution under RG flow. We study 
a couple of examples here.

Consider a fractional D1-brane in the twisted circle, defined as a brane wrapped on the cycle $S^1_A$.\foot{It turns out that this brane reduces to the fractional D0-brane 
in ${\bf C}/{\bf Z}_n$ upon taking the $R\to 0$ limit.} 
In the IR region, as we saw in the previous subsection, this cycle
completely disappears. Thus we conclude 
that the fractional D1-brane will disappear after the tachyon 
condensation. In other words, the D1-brane self-annihilates via open string tachyon condensation at the point $P_1$ where the
cycle becomes topologically trivial. On the other hand, a bulk D1-brane
corresponds to the brane wrapped on $S^1_C$, which will 
become a bulk D0-brane in the orbifold limit.
This cycle evolves into a regular circle in the IR. For non-zero values
of $\rho$ (i.e.\ except the origin of ${\bf C}$), it evolves
smoothly. Thus a bulk D1-brane in the twisted circle will become a
usual D1-brane in flat space. However, if we put the D1-brane at
the origin, the corresponding cycle will vanish only at $P_1=0$.
In this case we have two possibilities: a bulk D1-brane at
the origin will either self-annihilate or become a usual D1-brane.
It  is also possible for a D1-brane wrapped on a compactified circle 
to be created out of nothing (i.e.\ the cycle 
$S^1_B$) in the process of tachyon condensation. 
Finally, the flows we have described above may be superposed to produce 
more complicated flows, for example one in which a fractional brane disappears
while at the same time a bulk brane is created out of nothing.
See \MiTa\ for details of various other possible flows.

It turns out to be possible confirm some of these geometrically inspired
conjectures using a more rigorous approach. As we have reviewed above,
it is possible to obtain exact results on closed string
tachyon condensation by studying ${\cal N}=2$ supersymmetric
renormalization group flows in GLSMs.
Introducing D-branes into the story corresponds to putting the
GLSMs on the disk with appropriate boundary conditions and adding
relevant boundary degrees of freedom (Chan-Paton factors)
\refs{\BSFTBA,\HoIqVa,\Gob}. This uses the fact that we can realize
various D-branes by open string tachyon condensation (or
equally the relevant perturbation in the sense of subsection 2.3) on
higher dimensional D-branes. When the boundary conditions and
interactions in question also preserve ${\cal N}=2$ supersymmetry,
it is not difficult to construct RG flows that follow the
evolution of these boundary conditions to late Euclidean times.
Using this method, we can indeed obtain the same results as before, as well as corresponding ones in the ${\bf C}/{\bf Z}_n$ system.

\newsec{Other Directions and Open Problems}

Recent investigations of closed string tachyon condensation, some of which
we have reviewed in this lecture, have thrown up several interesting
questions that remain unanswered. One of the most intriguing of these
concerns a generalization of spacetime energy. As we
argued in section 2, open string RG flows proceed in the direction to lower
the $g$ function, which may be identified spacetime energy onshell. We
also reviewed (see \GuHeMiSc\ for details) that spacetime energy may also be shown
to decrease along a class of ``mild" bulk RG flows: those that leave the spacetime
asymptotics unchanged. However we now have several examples 
of closed string tachyon condensation (e.g.\ the twisted circle---see section 4) in which the asymptotic structure of spacetime is drastically modified as the tachyon condensation process proceeds. In such examples, the energies of initial and final states cannot be compared using the
usual notion of ADM energy. Nonetheless it certainly seems reasonable that 
there is some sense in which the IR fixed point should be assigned lower 
energy than the UV fixed point in these RG flows.

All this appears to suggest that our current understanding of gravitational 
energy can be generalized, and that the world-sheet string theory---in 
particular RG flows on the world-sheet---may be useful in 
understanding this generalization. It is possible that the 
Zamolodchikov theorem, appropriately applied, would allow for the 
identification of some quantity that decreases along RG flows, and that one may then regard as a generalization of ADM energy (see \GuHeMiSc\ for speculations on how this might work). See 
\refs{\Dabh,\chicago,\DhVa,\NamTS,\SinQS,\BasuJT, \KrausCB,\DavidKM,\ChaudhuriJJ,\SinXE} 
for related work. 

Relatedly, we would also like to point out the conjectured existence 
\DavidKM\ of flows that drastically alter spacetime asymptotics
purely within the supergravity approximation.
This suggests that a generalization of ADM energy allowing one
to compare the energies of spacetimes with different asymptotics may already
exist within general relativity. Concrete results on this
subject (perhaps along the lines of \LiuBX) would be fascinating.

Recent investigations of closed string tachyon condensation have focused
on localized tachyons, for the reasons reviewed in subsection 2.5. 
However the tachyon most familiar to all students
of string theory---the ground state of the bosonic string---is a
bulk, or delocalized, tachyon. It would be truly fascinating to determine
the endpoint (if it exists) of this tachyonic instability. 
It would certainly be nice to know if the bosonic string lies within 
superstring configuration space \DavidNQ---or whether it has been ordained to exist 
merely to supply warm up exercises for string theory students. For the case of
homogeneous tachyon condensation it may be possible to 
regard the time dependent process as a time-like Liouville theory, as 
in the previous open string case \wilsont. 
See \refs{\DaCunhaFM,\StromingerFN,\SchomerusVV}
for preliminary work in this direction. 

A question that is similar
in spirit concerns the endpoint of the instability of
the type 0 string. It has been suggested, utilizing the identification of
the type 0 string as a type II flux brane, that the endpoint of type
0 tachyon condensation is the ground state of the type II string
\refs{\CostaNW,\GuSt,\RussoTF}.
This general proposal also receives support from the analysis of the decay
of twisted circles in the $n \to \infty$ limit (see section 4 and
\DaGuHeMi).  However, at present it seems fair to say that a really convincing 
analysis of type 0
tachyon condensation is as yet lacking. 

One idea for getting around the Zamolodchikov $c$-theorem in trying to relate bulk tachyon condensation to world-sheet RG flow, is to consider a non-critical string theory
(see \refs{\KutasovPF,\HsuCM} for earlier discussions of related issues 
in $c\leq 1$ string theory).
Then we need a Liouville-like field and this will lead to a spontaneous
breaking of the rotational symmetry. 
See \refs{\BardakciGV,\BardakciUI,\BardakciVS,\BardakciUX,\BardakciAN,\MaggioreQR,\SuyamaXK,\DineCA,\SuyamaKY,\SuyamaXJ,\SuyamaAS,\KarczmarekPV,\JoannaAndy}
for other investigations of bulk tachyon condensation.


\nref\KachruED{
S.~Kachru, J.~Kumar and E.~Silverstein,
``Orientifolds, RG flows, and closed string tachyons,''
Class.\ Quant.\ Grav.\  {\bf 17}, 1139 (2000)
[arXiv:hep-th/9907038].
}

\nref\AdamsJB{
A.~Adams and E.~Silverstein,
``Closed string tachyons, AdS/CFT, and large N QCD,''
Phys.\ Rev.\ D {\bf 64}, 086001 (2001)
[arXiv:hep-th/0103220].
}

\nref\SuyamaBN{
T.~Suyama,
``Closed string tachyons in non-supersymmetric heterotic theories,''
JHEP {\bf 0108}, 037 (2001)
[arXiv:hep-th/0106079].
}

\nref\SuyamaNE{
T.~Suyama,
``Melvin background in heterotic theories,''
Nucl.\ Phys.\ B {\bf 621}, 235 (2002)
[arXiv:hep-th/0107116].
}

\nref\TseytlinQB{
A.~A.~Tseytlin,
``Magnetic backgrounds and tachyonic instabilities in closed string
theory,''
arXiv:hep-th/0108140.
}

\nref\UrangaDX{
A.~M.~Uranga,
``Wrapped fluxbranes,''
arXiv:hep-th/0108196.
}

\nref\DabholkarGZ{
A.~Dabholkar,
``On condensation of closed-string tachyons,''
Nucl.\ Phys.\ B {\bf 639}, 331 (2002)
[arXiv:hep-th/0109019].
}

\nref\RussoNA{
J.~G.~Russo and A.~A.~Tseytlin,
``Supersymmetric fluxbrane intersections and closed string tachyons,''
JHEP {\bf 0111}, 065 (2001)
[arXiv:hep-th/0110107].
}

\nref\ArmoniUW{
A.~Armoni and E.~Lopez,
``UV/IR mixing via closed strings and tachyonic instabilities,''
Nucl.\ Phys.\ B {\bf 632}, 240 (2002)
[arXiv:hep-th/0110113].
}

\nref\NatsuumeQT{
M.~Natsuume,
``The heterotic enhancon,''
Phys.\ Rev.\ D {\bf 65}, 086002 (2002)
[arXiv:hep-th/0111044].
}

\nref\AdamsNE{
A.~Adams and M.~Fabinger,
``Deconstructing noncommutativity with a giant fuzzy moose,''
JHEP {\bf 0204}, 006 (2002)
[arXiv:hep-th/0111079].
}

\nref\SuyamaIK{
T.~Suyama,
``Charged tachyons and gauge symmetry breaking,''
JHEP {\bf 0202}, 033 (2002)
[arXiv:hep-th/0112101].
}

\nref\HorowitzUH{
G.~T.~Horowitz and T.~Jacobson,
``Note on gauge theories on M/Gamma and the AdS/CFT correspondence,''
JHEP {\bf 0201}, 013 (2002)
[arXiv:hep-th/0112131].
}

\nref\BarbonDI{
J.~L.~F.~Barbon and E.~Rabinovici,
``Closed-string tachyons and the Hagedorn transition in AdS space,''
JHEP {\bf 0203}, 057 (2002)
[arXiv:hep-th/0112173].
}

\nref\DeAlwisKP{
S.~P.~De Alwis and A.~T.~Flournoy,
``Closed string tachyons and semi-classical instabilities,''
Phys.\ Rev.\ D {\bf 66}, 026005 (2002)
[arXiv:hep-th/0201185].
}

\nref\FontPQ{
A.~Font and A.~Hernandez,
``Non-supersymmetric orbifolds,''
Nucl.\ Phys.\ B {\bf 634}, 51 (2002)
[arXiv:hep-th/0202057].
}

\nref\NekrasovKF{
N.~A.~Nekrasov,
``Milne universe, tachyons, and quantum group,''
Surveys High Energ.\ Phys.\  {\bf 17}, 115 (2002)
[arXiv:hep-th/0203112].
}

\nref\RabadanZQ{
R.~Rabadan and J.~Simon,
 ``M-theory lift of brane-antibrane systems and localised closed string
tachyons,''
JHEP {\bf 0205}, 045 (2002)
[arXiv:hep-th/0203243].
}

\nref\UrangaAG{
A.~M.~Uranga,
``Localized instabilities at conifolds,''
arXiv:hep-th/0204079.
}

\nref\ItoyamaDD{
H.~Itoyama and S.~Nakamura,
``Off-shell crosscap state and orientifold planes with background
dilatons,''
Nucl.\ Phys.\ B {\bf 644}, 248 (2002)
[arXiv:hep-th/0205163].
}

\nref\TakayanagiPI{
T.~Takayanagi,
``Modular invariance of strings on pp-waves with RR-flux,''
JHEP {\bf 0212}, 022 (2002)
[arXiv:hep-th/0206010].
}

\nref\SarkarWB{
T.~Sarkar,
``Brane probes, toric geometry, and closed 
string tachyons,''
Nucl.\ Phys.\ B {\bf 648}, 497 (2003)
[arXiv:hep-th/0206109].
}

\nref\MartinecXQ{
E.~J.~Martinec and W.~McElgin,
``Exciting AdS orbifolds,''
JHEP {\bf 0210}, 050 (2002)
[arXiv:hep-th/0206175].
}

\nref\BorundaRA{
M.~Borunda, M.~Serone and M.~Trapletti,
 ``On the quantum stability of type IIB orbifolds and orientifolds with
Scherk-Schwarz SUSY breaking,''
Nucl.\ Phys.\ B {\bf 653}, 85 (2003)
[arXiv:hep-th/0210075].
}

\nref\BarbonNW{
J.~L.~F.~Barbon and E.~Rabinovici,
``Remarks on black hole instabilities and closed string tachyons,''
Found.\ Phys.\  {\bf 33}, 145 (2003)
[arXiv:hep-th/0211212].
}

\nref\KatzJH{
S.~Katz, T.~Pantev and E.~Sharpe,
``D-branes, orbifolds, and Ext groups,''
Nucl.\ Phys.\ B {\bf 673}, 263 (2003)
[arXiv:hep-th/0212218].
}

\nref\ArmoniVA{
A.~Armoni, E.~Lopez and A.~M.~Uranga,
``Closed strings tachyons and non-commutative instabilities,''
JHEP {\bf 0302}, 020 (2003)
[arXiv:hep-th/0301099].
}

\nref\HeYW{
Y.~H.~He,
``Closed string tachyons, non-supersymmetric orbifolds and generalised
McKay
correspondence,''
Adv.\ Theor.\ Math.\ Phys.\  {\bf 7}, 121 (2003)
[arXiv:hep-th/0301162].
}

\nref\NakamuraWZ{
S.~Nakamura,
 ``Decay rates of fixed planes and closed-string tachyons on unstable
orbifolds,''
arXiv:hep-th/0305054.
}

\nref\BakasEU{
I.~Bakas,
``Renormalization group flows and continual Lie algebras,''
JHEP {\bf 0308}, 013 (2003)
[arXiv:hep-th/0307154].
}

\nref\PiolineBS{
B.~Pioline and M.~Berkooz,
``Strings in an electric field, and the Milne universe,''
JCAP {\bf 0311}, 007 (2003)
[arXiv:hep-th/0307280].
}

\nref\SinYM{
S.~J.~Sin,
``Comments on the fate of unstable orbifolds,''
Phys.\ Lett.\ B {\bf 578}, 215 (2004)
[arXiv:hep-th/0308028].
}

\nref\LeeAR{
S.~g.~Lee and S.~J.~Sin,
``Chiral rings and GSO projection in orbifolds,''
Phys.\ Rev.\ D {\bf 69}, 026003 (2004)
[arXiv:hep-th/0308029].
}

\nref\ImamuraCX{
Y.~Imamura,
``Decay of type 0 NS5-branes to nothing,''
Phys.\ Rev.\ D {\bf 69}, 026005 (2004)
[arXiv:hep-th/0309024].
}

\nref\GrootNibbelinkZJ{
S.~Groot Nibbelink and M.~Laidlaw,
 ``Stringy profiles of gauge field tadpoles 
near orbifold singularities.
I:
Heterotic string calculations,''
JHEP {\bf 0401}, 004 (2004)
[arXiv:hep-th/0311013].
}

\nref\McInnesYA{
B.~McInnes,
``Orbifold physics and de Sitter spacetime,''
arXiv:hep-th/0311055.
}

\nref\MoellerGG{
N.~Moeller and M.~Schnabl,
``Tachyon condensation in open-closed p-adic string theory,''
JHEP {\bf 0401}, 011 (2004)
[arXiv:hep-th/0304213].
}

\nref\AndreevHN{
O.~Andreev,
``Comments on tachyon potentials in closed and open-closed string  theories,''
Nucl.\ Phys.\ B {\bf 680}, 3 (2004)
[arXiv:hep-th/0308123].
}

\nref\GarousiDB{
M.~R.~Garousi,
``S-matrix elements and closed string tachyon couplings in type 0 theory,''
arXiv:hep-th/0309028.
}

\nref\BakasXW{
I.~Bakas,
``Ricci flows and infinite dimensional algebras,''
arXiv:hep-th/0312274.
}


The reader will find several other interesting puzzles, questions and directions that we do not have the time to discuss, in a number of recent publications \refs{\KachruED - \BakasXW}.

\vskip 0.4 in

\centerline{{\bf Acknowledgements}}

We would like to thank all of our collaborators and several of our colleagues for many useful discussions over the span of several years on the topics reviewed in these lectures. We would also like to thank A. Adams, A. Dabholkar, Y. Okawa, and especially B. Zwiebach for useful comments on the manuscript. The work of M.H. was supported by a Pappalardo Fellowship from MIT, and by funds provided by the U.S. Department of Energy under cooperative research agreement DF-FC02-94ER40818. The work of S.M. was supported in part by DOE grant DE-FG03-91ER40654, in part by NSF grant number PHY-0239626, and in part by a Sloan Fellowship. The work of T.T.  was supported in part by DOE grant DE-FG03-91ER40654.

\appendix{A}{Spectrum of ${\bf C}/{\bf Z}_n$ Orbifold}

As discussed in subsection 3.1, the boundary conditions \tbc\ in the twisted sectors of the ${\bf C}/{\bf Z}_n$ orbifold shift the modings of the $Z$, $\psi$, and $\tilde\psi$ oscillators. One result, as we will see below, is that the world-sheet fermion number $Q$, which is an integer for superstrings in flat space, is instead valued in ${\bf Z}+k/n$, while $\tilde Q$ is valued in ${\bf Z}-k/n$.\foot{$Q$ and $\tilde Q$ are defined as the $H$ and $\tilde H$ momenta in the bosonized theories. They are also the contributions of the $\psi$ and $\tilde\psi$ theories to the spacetime angular momentum $J$ in the 8--9 plane. We are following the notations and conventions of \PolchinskiRR.} The $+$ and $-$ GSO projections, normally $e^{\pi iQ}=\pm1$ and $e^{\pi i\tilde Q}=\pm1$, must correspondingly be adjusted; they become
\eqn\gsos{
e^{\pi i(Q+k(1-1/n))} = \pm1, \qquad e^{\pi i(\tilde Q+k(1+1/n))} = \pm1.
}
With this adjustment, the usual type IIA/B sectors, NS$+$ and R$+$ on the left and NS$+$ and R$\mp$ on the right, yield orbifold theories with consistent interactions and modular-invariant torus partition functions. The exponents in \gsos\ involve $k(1\pm1/n)$ rather than simply $\pm k/n$ in order to make the left-hand side well defined, given that $k$ is only defined mod $n$. As we will see below, the extra $e^{\pi ik}$ is also related by modular invariance to the $(-1)^F$ appearing in the orbifold generator \orbgen, and in turn accounts for the ``reversed" GSO projection in the odd twisted sectors.

We will write the torus partition function for the type IIB orbifold, which is slightly simpler since the left- and right-moving GSO projections are the same; the generalization to IIA should be clear. It is:
\eqn\toruspartitionfunction{
Z_{T^2} = iL^2\int_F{d^2\tau\over16\pi^2\alpha'\tau_2^2}Z_{\perp},
}
where $L$ is an IR length scale used to regulate the zero-modes of the string, $F$ is the modular domain, and
\eqn\Zdef{\eqalign{
Z_{\perp} &= 
Z_X^6{1\over n}\sum_{k,k'}Z_Z^{(k,k')}Z_{\psi^\perp}^{(k,k')}Z_{\tilde\psi^\perp}^{(k,k')}, \cr
Z_X &= L(4\pi^2\alpha')^{-1/2}|\eta(\tau)|^{-2}, \cr
Z_Z^{(k,k')} &= 
\cases{ Z_X^2, & $(k,k') = (0,0)$ \cr \left|Z_{1+2k'/n}^{1+2k/n}\right|^{-2} & $(k,k') \neq (0,0)$ }, \cr
Z_{\psi^\perp}^{(k,k')} &= {1\over2}\left[
\left(Z_0^0\right)^3Z_{2k'/n}^{2k/n} 
- (-1)^{k'}\left(Z_0^1\right)^3Z_{2k'/n}^{1+2k/n} \right. \cr
&\quad\qquad\qquad\qquad\,\left. - e^{\pi ik(1-1/n)}\left(Z_1^0\right)^3Z_{1+2k'/n}^{2k/n}
-(-1)^{k'}e^{\pi ik(1-1/n)}\left(Z_1^1\right)^3Z_{1+2k'/n}^{1+2k/n}
\right], \cr
Z_{\tilde\psi^\perp}^{(k,k')} &= {Z_{\psi^\perp}^{(k,k')}}^*, \cr
Z_\beta^\alpha &= {1\over\eta(\tau)}\vartheta\left[{\alpha/2\atop\beta/2}\right](0,\tau).
}}
The sum over $k$ and $k'$ may be taken over any set of representatives of ${\bf Z}_n$, since the summand is invariant under $k\to k+n$ as well as $k'\to k'+n$. As usual, $k$ labels the twisted sector. The sum over $k'$ enforces the orbifold projection; each term has the operator $(-1)^{Fk'}e^{2\pi iJk'/n}$ inserted into the path integral. In the expression for $Z_{\psi^\perp}^{(k,k')}$, the second and fourth terms represent R states, and their $(-1)^{k'}$ coefficients are from this $(-1)^{Fk'}$. The modular transformation $\tau\to-1/\tau$ relates these coefficients to the factor $e^{\pi ik}$ appearing in the coefficients of the third and fourth terms, which in turn define the GSO projections \gsos.\foot{In checking the invariance of this partition function under the modular transformation $\tau\to-1/\tau$, the following identity (which corrects equation (10.7.14) of \PolchinskiRR) is useful:
\eqn\polcorrect{
Z^\alpha_\beta(\tau) = e^{\pi i\alpha\beta/2}Z^\beta_{-\alpha}(-1/\tau).
}
}

In discussing the spectrum, it is useful, as we did in subsection 3.1, to choose the twist number $k$ to lie in the range $-n/2<k<n/2$. The NS ground state of the $\psi$ field has world-sheet fermion number $Q_\psi$ equal to $k/n$. Including the ghost contribution, the ground state has total fermion number
\eqn\nsq{
Q|0\lb_{\rm NS} = \left(1+{k\over n}\right)|0\lb_{\rm NS}.
}
Since the fermionic operators can raise and lower $Q$ only by integers, we see that it is indeed valued in ${\bf Z}+k/n$. Similarly, on the right-moving side the ground state has $\tilde Q=1-k/n$, and $\tilde Q\in {\bf Z}-k/n$ for all states. We also see that the $+$ GSO projection \gsos\ projects the ground state in for odd $k$ and out for even $k$, as claimed in subsection 3.1.

Under the usual R periodicity ($k=0$), the $\psi$ theory has two degenerate ground states, with $Q_\psi=\pm{1\over2}$. At non-zero $k$, the shifted moding breaks the degeneracy, leading to a single ground state with $Q_\psi$ equal to ${1\over2}+k/n$ ($k<0$) or $-{1\over2}+k/n$ ($k>0$). Including the other 8 fermions and the ghosts, we find a 16-fold degenerate ground state with $Q\in{\bf Z}+k/n$. Half of the ground states are projected in by either GSO projection \gsos. The analysis is similar for the right-movers.

\listrefs

\end